\documentclass[12pt]{article}  
\usepackage{cite}
\usepackage{epsfig}
\usepackage{graphicx}
\usepackage{amsmath}
\usepackage{amssymb}
\usepackage{ulem}
\usepackage{mdwlist} 
\usepackage{color}  

\usepackage{a41}
\usepackage{color}
\usepackage[rflt]{floatflt}
\usepackage{float}
\usepackage{slashed}

\usepackage{array}

\usepackage[english]{babel}
\usepackage[latin1]{inputenc}
\usepackage[T1]{fontenc}
\usepackage{ae}

\usepackage{url}

\usepackage{amsmath, amsthm, amssymb}
\newtheorem{thm}{Theorem}[section]

\newtheorem{definition}[thm]{Definition}

 \newcommand{\GeV}{\mathrm{GeV}}

 \newcommand{\MS}{\overline{\sf MS}}

 \newcommand{\Athathat}{\hat{\hspace*{0mm}\hat{\tilde{A}}}}

\newcommand{\NN}{\nonumber}

\newcommand{\ep}{\varepsilon}

\usepackage{rotating}

\usepackage{graphicx}

\newcounter{mmacnt}
\def\restartmma{\setcounter{mmacnt}{0}}
\restartmma \catcode`|=\active
\def|#1|{\mathrm{#1}}
\catcode`|=12
\newenvironment{mma}{
 \par\smallskip
 \catcode`|=\active
 \parskip=0pt\parindent=0pt 
 \small
 \def\In##1\\{%
\def\linebreak{\hfill\break\null\qquad}%
\refstepcounter{mmacnt}
\hangindent=2.5em\hangafter=0
\leavevmode
\llap{\tiny\sffamily n[\arabic{mmacnt}]:=\kern.5em}%
\mathversion{bold}\footnotesize$\displaystyle##1$\normalsize
\mathversion{normal}\par
 }%
 \def\Print##1\\{%
\def\linebreak{\hfill\break}%
\hangindent=2.5em\hangafter=0
\leavevmode ##1\par}%
 \def\Out##1\\{%
\def\linebreak{$\hfill\break\null\hfill$}%
\kern\abovedisplayskip\par
\hangindent=2.5em\hangafter=0
\leavevmode
\llap{\tiny\sffamily Out[\arabic{mmacnt}]=\kern.5em}
\footnotesize$\displaystyle##1$\normalsize\hfill\null\par
\kern\belowdisplayskip
 }%
 \def\Warning##1##2\\{%
\def\linebreak{\hfill\break}%
\hangindent=2.5em\hangafter=0
\leavevmode
{\scriptsize##1 : ##2}\par}%
}{%
 \par\smallskip
}

\usepackage{color}

\newenvironment{fshaded}{%
\MakeFramed {\FrameRestore}
}%
{\endMakeFramed}

\allowdisplaybreaks[4]

\begin{document}
\setlength{\baselineskip}{0.515cm}
\sloppy
\thispagestyle{empty}
\begin{flushleft}
DESY 17--195
\\
DO--TH 17/32\\
November  2017\\
\end{flushleft}

\mbox{}
\vspace*{\fill}
\begin{center}

{\LARGE\bf The two-mass contribution to the three-loop}

\vspace*{3mm} 
{\LARGE\bf pure singlet operator matrix element}

\vspace{3cm}
\large
J.~Ablinger$^a$, 
J.~Bl\"umlein$^b$, 
A.~De Freitas$^b$, 
C.~Schneider$^a$, \\  and  K.~Sch\"onwald$^b$ 

\vspace{1.cm}
\normalsize
{\it $^a$~Research Institute for Symbolic Computation (RISC),\\
  Johannes Kepler University, Altenbergerstra{\ss}e 69,
  A--4040, Linz, Austria}\\

\vspace*{3mm}
{\it  $^b$ Deutsches Elektronen--Synchrotron, DESY,}\\
{\it  Platanenallee 6, D-15738 Zeuthen, Germany}
\\

\end{center}
\normalsize
\vspace{\fill}
\begin{abstract}
\noindent
We present the two-mass QCD contributions to the pure singlet operator matrix element at three loop order 
in $x$-space. These terms are relevant for calculating the structure function $F_2(x,Q^2)$ at $O(\alpha_s^3)$
as well as for the matching relations in the variable flavor number scheme and the heavy quark distribution
functions at the same order. The result for the operator matrix element is given in terms of generalized 
iterated integrals that include square root letters in the alphabet, depending also on the mass ratio through 
the main argument. Numerical results are presented.
\end{abstract}

\vspace*{\fill}
\noindent
\numberwithin{equation}{section}
\newpage 
\section{Introduction}
\label{sec:1}

\vspace*{1mm}
\noindent
Starting at 2-loop order, massive operator matrix elements (OME) $A_{ij}$, which are the transition matrix elements
in the variable flavor number scheme (VFNS), receive two-mass contributions \cite{VFNS,Ablinger:2017err}. This also 
applies to the Wilson coefficients in deeply inelastic scattering. The single mass contributions to the pure 
singlet ({\sf PS}) OME, $A_{Qq}^{{\sf PS},(3)}$, have been calculated in Ref.~\cite{Ablinger:2014nga}. In the present 
paper we 
present the corresponding 2-mass contributions. They occur first at 3--loop order. Previously, we have calculated 
already the fixed moments of this 
OME for the Mellin variable $N = 2,4,6$, to $O(\eta^3 \ln^3(\eta)$, with $\eta < 1$ the mass ratio of the heavy
quarks squared in \cite{MOM,Ablinger:2017err} using the package {\tt Q2E} 
\cite{Harlander:1997zb,Seidensticker:1999bb}, as well as the two-mass contributions 
to the 3-loop OMEs in the flavor non-singlet cases $A_{qq,Q}^{{\sf NS},(3)}$ and $A_{qq,Q}^{{\sf NS-TR},(3)}$ and 
for $A_{gq,Q}^{(3)}$, cf. \cite{Ablinger:2017err}, both in $N$- and in $x$-space. The 2-mass contributions to 
$A_{gg,Q}^{(3)}$ are in preparation 
\cite{AGG2}. Various other 3-loop single mass contributions have also been completed, cf.~\cite{Ablinger:2010ty,
Blumlein:2012vq,
Behring:2015roa,
Behring:2015zaa,
Ablinger:2014lka,
Ablinger:2014uka,
Ablinger:2014vwa,
Behring:2016hpa,
Blumlein:2016xcy,AGG} and all 
the logarithmic contributions are known to this order \cite{Behring:2014eya}.
Furthermore, for the OME $A_{Qg}^{(3)}$, all 
diagrams consisting of contributions that can be obtained in terms of first order factorizing 
differential or 
difference equations have been calculated \cite{Ablinger:2015tua,Ablinger:2016swq,PROC1}. For all OMEs a 
series of moments has been calculated in the single mass case in Ref.~\cite{Bierenbaum:2009mv} 
using the code {\tt MATAD} \cite{Steinhauser:2000ry}. From the single pole terms of 
all the OMEs, we have derived the contributing 3-loop anomalous dimensions, cf. 
e.g.~\cite{Ablinger:2014nga,Ablinger:2017tan}.

We perform the calculation of the 2-mass part of the OME $A_{Qq}^{{\sf PS},(3)}$ mainly in $x$-space\footnote{This 
strategy 
has also been used in Ref.~\cite{Blumlein:2011mi}.}, using only some elements of 
the $N$-space formalism. In the presnt case one obtains first 
order factorizable expressions in $x$-space, but not in $N$-space. This implies that the $N$-space solution cannot be 
given by sum and product expressions only. 

The paper is organized as follows. In Section~\ref{sec:2} we present the renormalized pure-singlet OME in the 2-mass 
case. Details of the calculation are given in Section~\ref{sec:3}. In Section~\ref{sec:4} the result of the 
calculation is presented and numerical results are given in Section~\ref{sec:5}. Section~\ref{sec:6} contains the 
conclusions. Details on new iterated integrals emerging in the representation, the $G$-functions, and a series of 
fixed moments as a function of the mass ratio of the two heavy quarks for the unrenormalized OME are given in the Appendix.
\section{The renormalized 2-mass pure singlet OME}
\label{sec:2}

\vspace*{1mm}
\noindent
The generic pole structure for the {\sf PS} three--loop two--mass contribution is given by 
\cite{Ablinger:2017err}
\begin{eqnarray}
\Athathat_{Qq}^{(3),\rm{PS}} &=&
\frac{8}{3 \ep^3} \gamma_{gq}^{(0)} \hat{\gamma}_{qg}^{(0)} \beta_{0,Q}
+\frac{1}{\ep^2}\biggl[
2 \gamma_{gq}^{(0)} \hat{\gamma}_{qg}^{(0)} \beta_{0,Q} \left(L_1+L_2\right)
+\frac{1}{6} \hat{\gamma}_{qg}^{(0)} \hat{\gamma}_{gq}^{(1)}
-\frac{4}{3} \beta_{0,Q} \hat{\gamma}_{qq}^{\rm{PS},(1)}
\biggr] 
\NN\\&&
+\frac{1}{\ep}
\biggl[
\gamma_{gq}^{(0)} \hat{\gamma}_{qg}^{(0)} \beta_{0,Q} \left(L_1^2+L_1 L_2+L_2^2\right)
+\biggl\{\frac{1}{8} \hat{\gamma}_{qg}^{(0)} \hat{\gamma}_{gq}^{(1)}
- \beta_{0,Q} \hat{\gamma}_{qq}^{\rm{PS},(1)}
\biggr\} \left(L_2+L_1\right)
\NN\\&&
+\frac{1}{3} \hat{\tilde{\gamma}}_{qq}^{(2),\rm{PS}}
-8 a_{Qq}^{(2),\rm{PS}} \beta_{0,Q}
+ \hat{\gamma}_{qg}^{(0)} a_{gq}^{(2)}
\biggr] 
+\tilde{a}_{Qq}^{(3),\rm{PS}}\left(m_1^2,m_2^2,\mu^2\right)
                   \label{Ahhhqq3PSQ},
\end{eqnarray}
where we used the short hand notation\footnote{In Eqs.~(3.110, 3.111) of \cite{Ablinger:2017err} unfortunately 
only the shift $N_F + 1 \rightarrow N_F$ has been used, which we correct here.}
\begin{eqnarray}
\hat{\gamma}_{ij} &=& \gamma_{ij}(N_F+2) - \gamma_{ij}(N_F)
\\
\hat{\tilde{\gamma}}_{ij} &=& \frac{\gamma_{ij}(N_F+2)}{N_F+2}  - \frac{\gamma_{ij}(N_F)}{N_F}.
\end{eqnarray}
The tilde in $\Athathat_{Qq}^{(3),\rm{PS}}$ indicates that we are considering only the genuine two-mass contributions, and the double hat is used
to denote a completely unrenormalized OME. Here the $\gamma_{ij}^{(l)}$'s are anomalous dimensions at $l+1$ loops, $\beta_{0,Q}=-\frac{4}{3} T_F$, and
\begin{equation}
L_1 = \ln\left(\frac{m_1^2}{\mu^2}\right), \quad L_2 = \ln\left(\frac{m_2^2}{\mu^2}\right),
\label{L1L2}
\end{equation}
where $m_1$ and $m_2$ are the masses of the heavy quarks, and $\mu$ is the renormalization scale. Our goal is to compute the $O(\ep^0)$ term $\tilde{a}_{Qq}^{(3),\rm{PS}}\left(m_1^2,m_2^2,\mu^2\right)$.

In the ${\MS}$--scheme, renormalizing the heavy masses on-shell, the renormalized expression is given by
\begin{eqnarray}
\tilde{A}_{Qq}^{(3), \MS, \rm{PS}} &=&
-\gamma_{gq}^{(0)} \hat{\gamma}_{qg}^{(0)} \beta_{0,Q} 
\left(\frac{1}{4} L_2^2 L_1+ \frac{1}{4} L_1^2 L_2+ \frac{1}{3} L_1^3+\frac{1}{3} L_2^3\right)
\NN\\&&
+\biggl\{
-\frac{1}{16} \hat{\gamma}_{qg}^{(0)} \hat{\gamma}_{gq}^{(1)}
+ \frac{1}{2}\beta_{0,Q} \hat{\gamma}_{qq}^{\rm{PS},(1)}
\biggr\}
\left(L_2^2+L_1^2\right)
\NN\\&&
+\biggl\{
4 a_{Qq}^{(2), \rm{PS}} \beta_{0,Q}
- \frac{1}{2} \hat{\gamma}_{qg}^{(0)} a_{gq}^{(2)}
-\frac{1}{4} \beta_{0,Q} \zeta_2 \gamma_{gq}^{(0)} \hat{\gamma}_{qg}^{(0)}
\biggr\} \left(L_1+L_2\right)
\NN\\&&
+8 \overline{a}_{Qq}^{(2), \rm{PS}} \beta_{0,Q}
- \hat{\gamma}_{qg}^{(0)} \overline{a}_{gq}^{(2)}
+\tilde{a}_{Qq}^{(3), \rm{PS}}\left(m_1^2,m_2^2,\mu^2\right)
\label{Aqq3PSQMSren}.
\end{eqnarray}
The transition relations for the renormalization of the heavy quarks in the $\overline{\sf MS}$-scheme
is given in \cite{Ablinger:2014nga}, Eq.~(5.100), but it only applies to the equal mass case since for the unequal mass
case the first contributions emerge at 3--loop order. In Eqs. (\ref{Ahhhqq3PSQ}) and (\ref{Aqq3PSQMSren}), 
$a_{Qq}^{(2), \rm{PS}}$ and $a_{gq}^{(2)}$ represent the $O(\ep^0)$ terms
of the two-loop OMEs $\hat{\hat{A}}_{Qq}^{(2), \rm{PS}}$ and  $\hat{\hat{A}}_{gq}^{(2)}$, respectively, 
while $\overline{a}_{Qq}^{(2), \rm{PS}}$ and $\overline{a}_{gq}^{(2)}$ represent the corresponding $O(\ep)$ terms,
cf.~Refs.~\cite{Buza:1995ie,Buza:1996wv,Bierenbaum:2007qe,Bierenbaum:2008yu,Bierenbaum:2009zt}.
Here and in what follows, $\zeta_k = \zeta(k),~k \in \mathbb{N}, k \geq 2$ denotes the Riemann $\zeta$-function.
\section{Details of the calculation}
\label{sec:3}

\vspace*{1mm}
\noindent
\subsection{The basic formalism}
\label{sec:31}

\begin{figure}[t]
\begin{center}
\begin{minipage}[c]{0.19\linewidth}
     \includegraphics[width=1\textwidth]{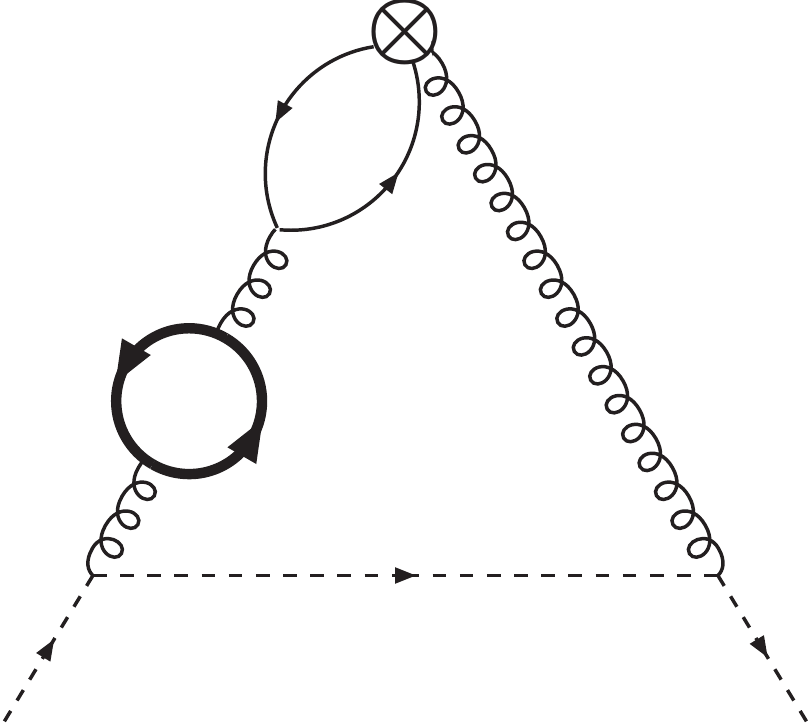}
\vspace*{-11mm}
\begin{center}
{\footnotesize (1)}
\end{center}
\end{minipage}
\hspace*{1mm}
\begin{minipage}[c]{0.19\linewidth}
     \includegraphics[width=1\textwidth]{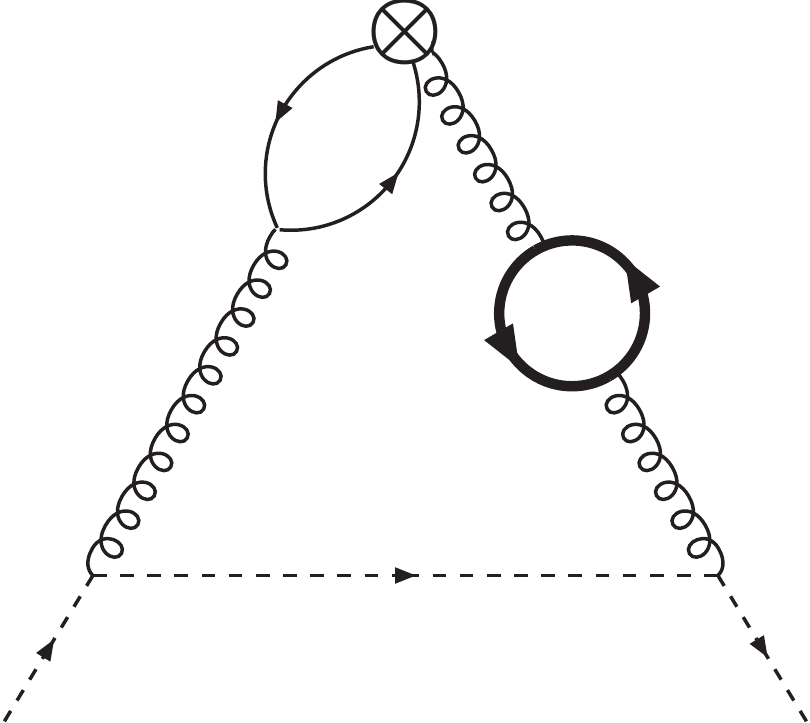}
\vspace*{-11mm}
\begin{center}
{\footnotesize (2)}
\end{center}
\end{minipage}
\hspace*{1mm}
\begin{minipage}[c]{0.19\linewidth}
     \includegraphics[width=1\textwidth]{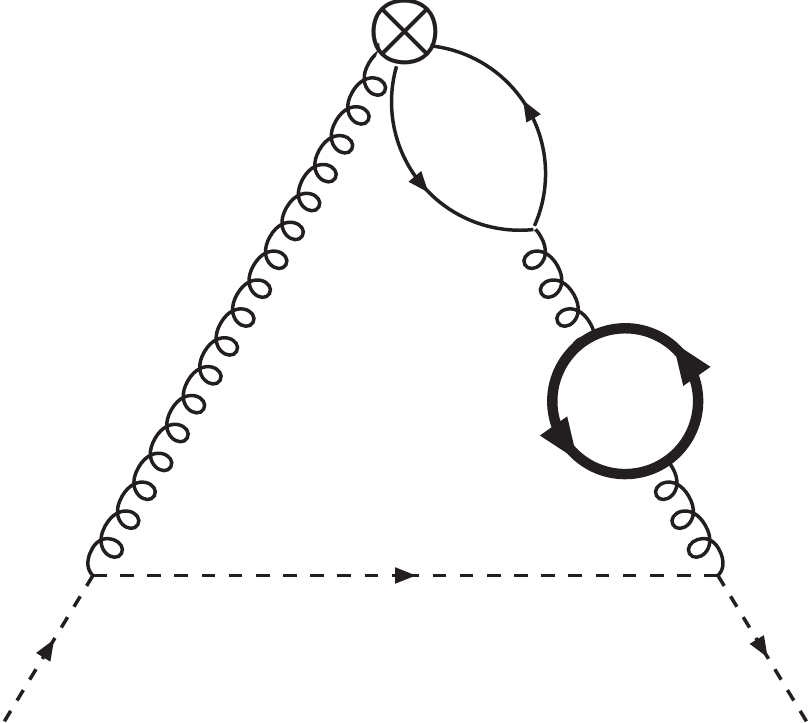}
\vspace*{-11mm}
\begin{center}
{\footnotesize (3)}
\end{center}
\end{minipage}
\hspace*{1mm}
\begin{minipage}[c]{0.19\linewidth}
     \includegraphics[width=1\textwidth]{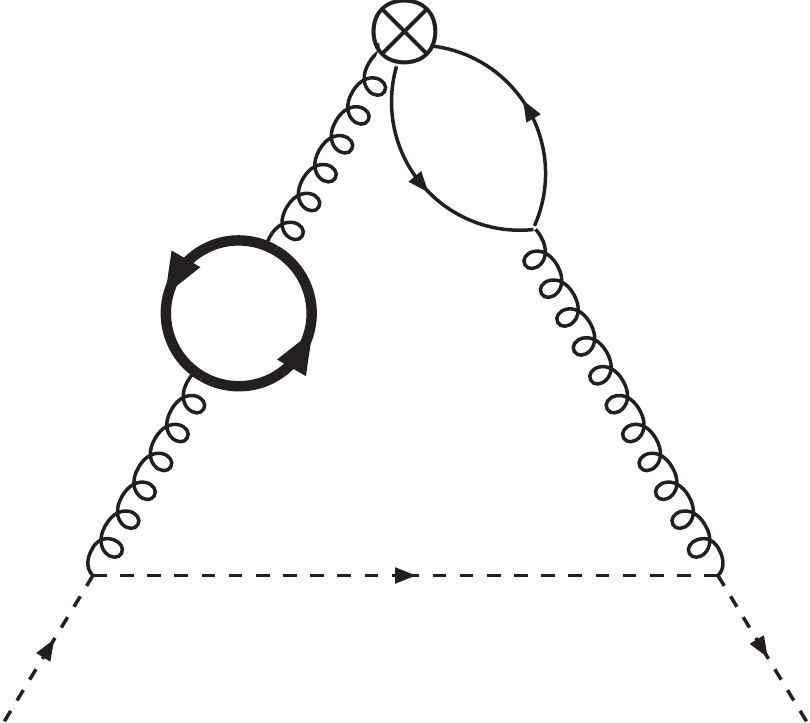}
\vspace*{-11mm}
\begin{center}
{\footnotesize (4)}
\end{center}
\end{minipage}

\vspace*{5mm}

\begin{minipage}[c]{0.19\linewidth}
     \includegraphics[width=1\textwidth]{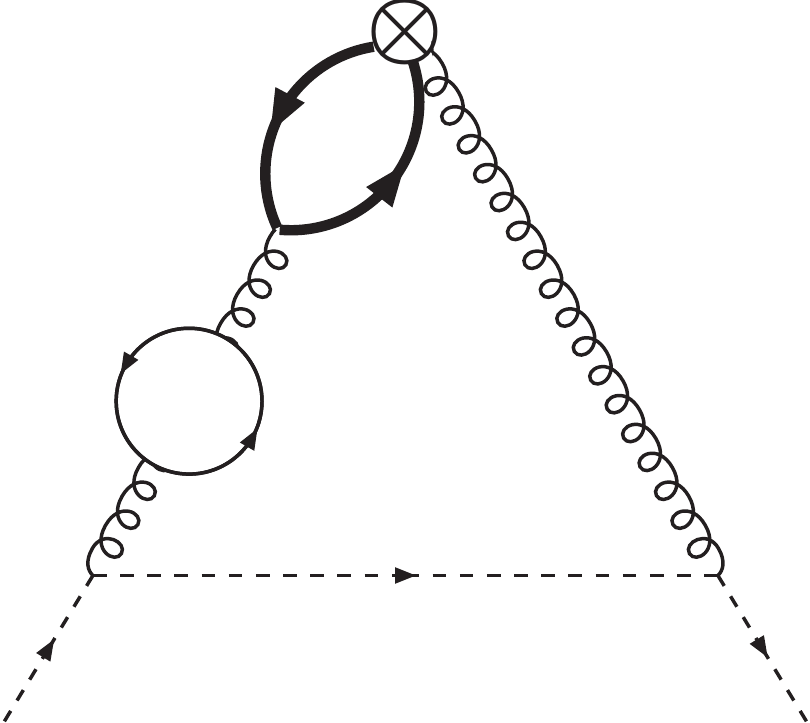}
\vspace*{-11mm}
\begin{center}
{\footnotesize (5)}
\end{center}
\end{minipage}
\hspace*{1mm}
\begin{minipage}[c]{0.19\linewidth}
     \includegraphics[width=1\textwidth]{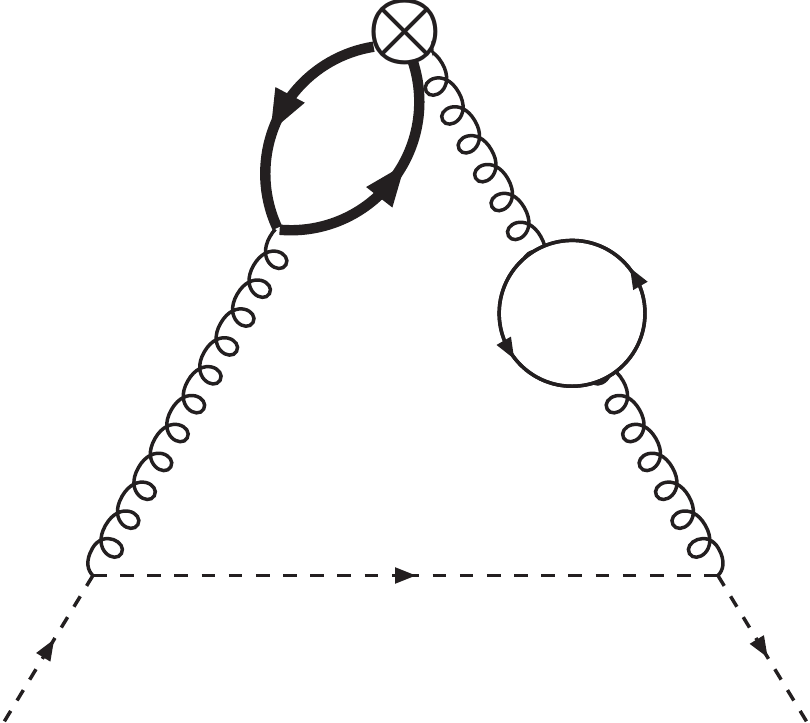}
\vspace*{-11mm}
\begin{center}
{\footnotesize (6)}
\end{center}
\end{minipage}
\hspace*{1mm}
\begin{minipage}[c]{0.19\linewidth}
     \includegraphics[width=1\textwidth]{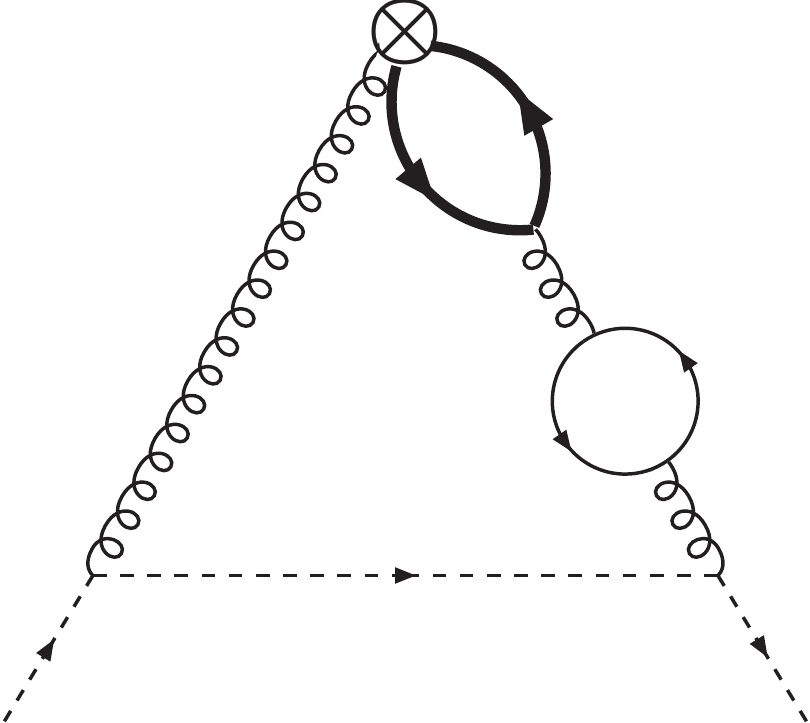}
\vspace*{-11mm}
\begin{center}
{\footnotesize (7)}
\end{center}
\end{minipage}
\hspace*{1mm}
\begin{minipage}[c]{0.19\linewidth}
     \includegraphics[width=1\textwidth]{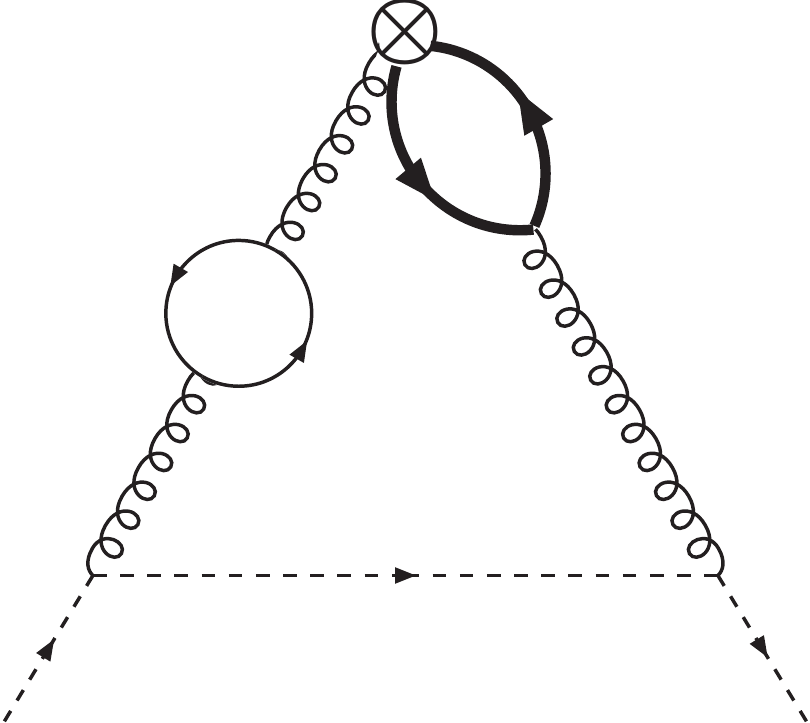}
\vspace*{-11mm}
\begin{center}
{\footnotesize (8)}
\end{center}
\end{minipage}

\vspace*{5mm}

\begin{minipage}[c]{0.19\linewidth}
     \includegraphics[width=1\textwidth]{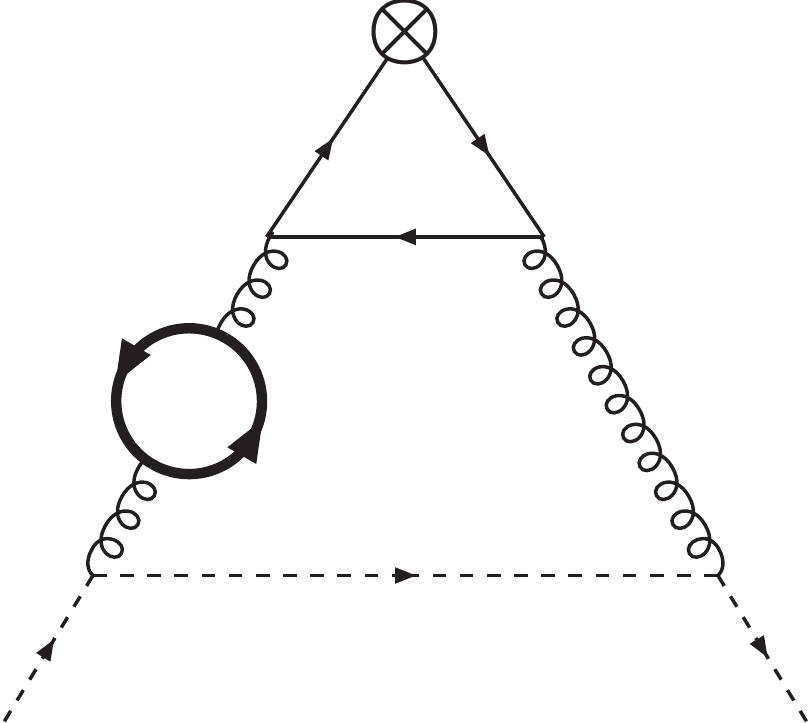}
\vspace*{-11mm}
\begin{center}
{\footnotesize (9)}
\end{center}
\end{minipage}
\hspace*{1mm}
\begin{minipage}[c]{0.19\linewidth}
     \includegraphics[width=1\textwidth]{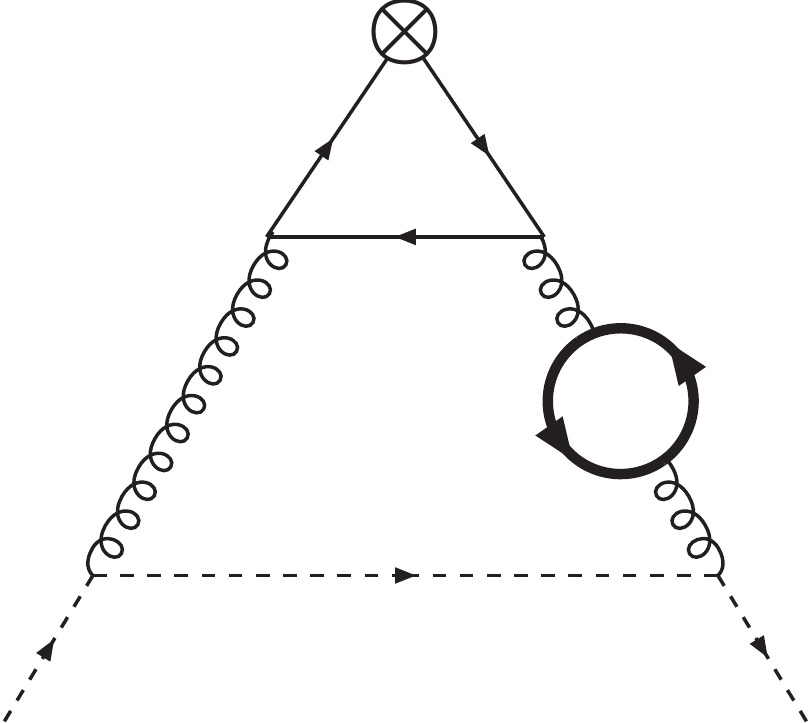}
\vspace*{-11mm}
\begin{center}
{\footnotesize (10)}
\end{center}
\end{minipage}
\hspace*{1mm}
\begin{minipage}[c]{0.19\linewidth}
     \includegraphics[width=1\textwidth]{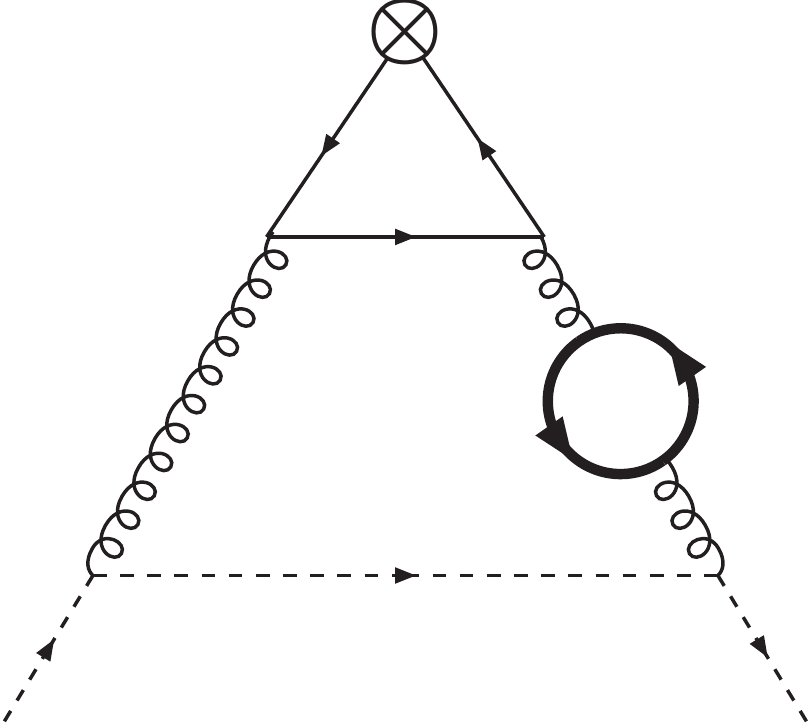}
\vspace*{-11mm}
\begin{center}
{\footnotesize (11)}
\end{center}
\end{minipage}
\hspace*{1mm}
\begin{minipage}[c]{0.19\linewidth}
     \includegraphics[width=1\textwidth]{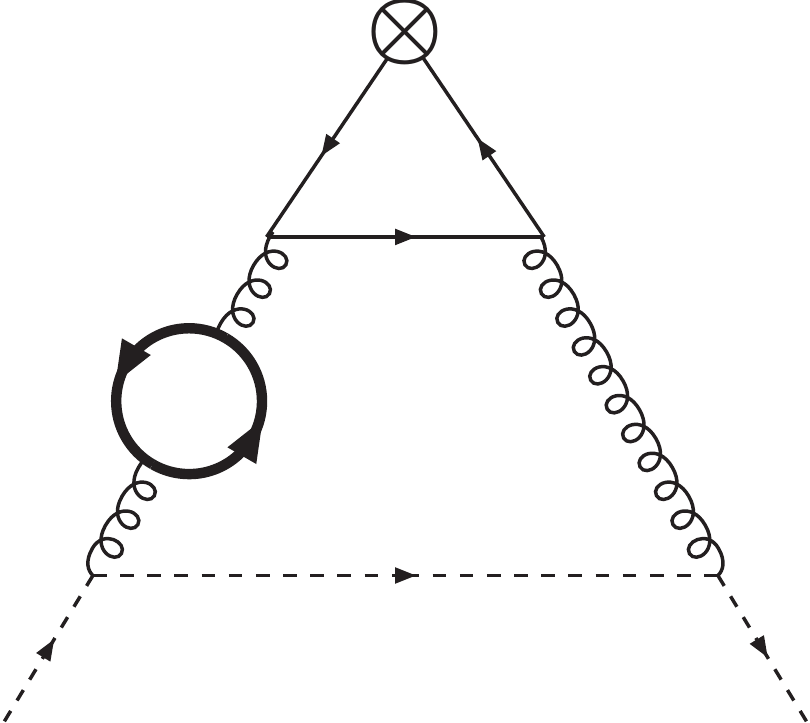}
\vspace*{-11mm}
\begin{center}
{\footnotesize (12)}
\end{center}
\end{minipage}

\vspace*{5mm}

\begin{minipage}[c]{0.19\linewidth}
     \includegraphics[width=1\textwidth]{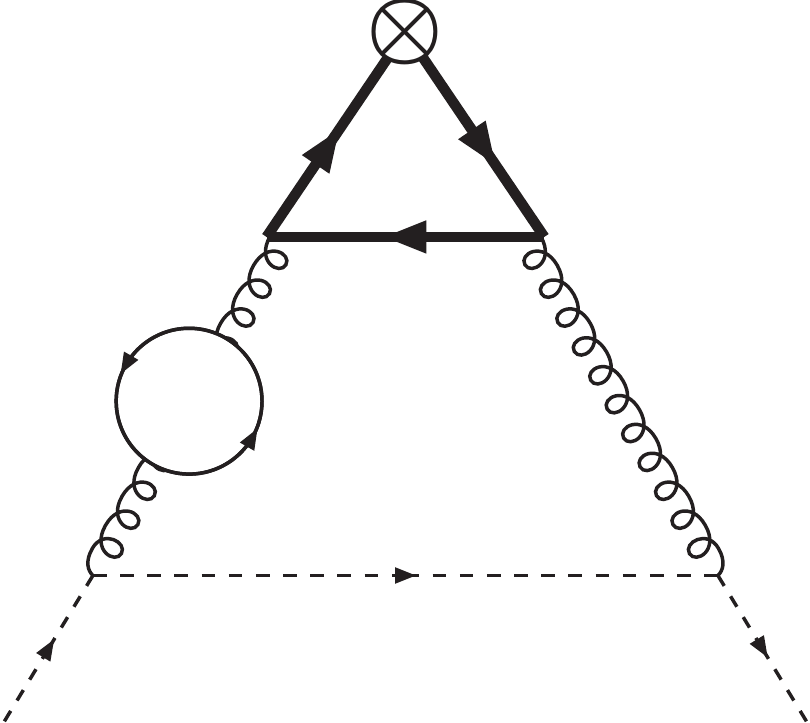}
\vspace*{-11mm}
\begin{center}
{\footnotesize (13)}
\end{center}
\end{minipage}
\hspace*{1mm}
\begin{minipage}[c]{0.19\linewidth}
     \includegraphics[width=1\textwidth]{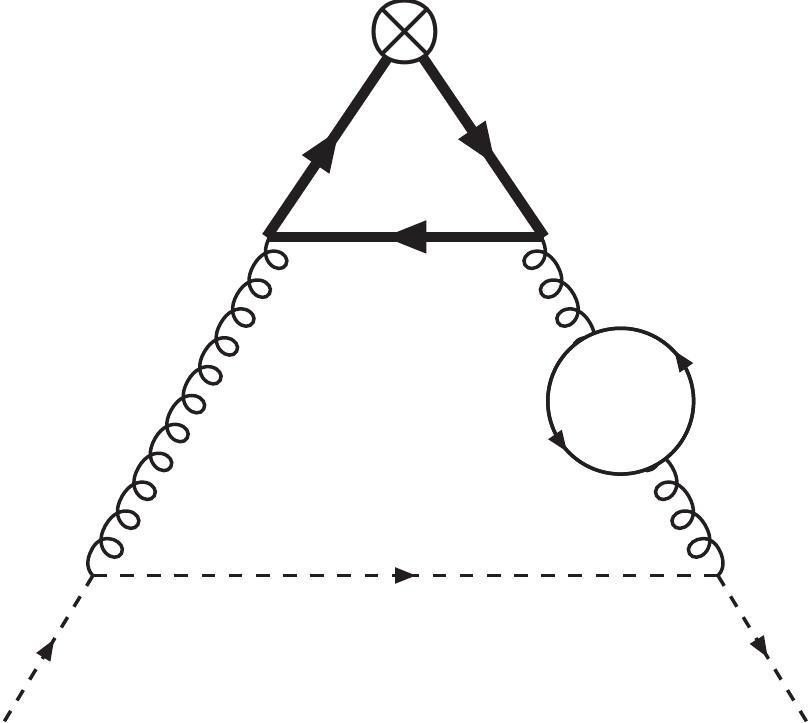}
\vspace*{-11mm}
\begin{center}
{\footnotesize (14)}
\end{center}
\end{minipage}
\hspace*{1mm}
\begin{minipage}[c]{0.19\linewidth}
     \includegraphics[width=1\textwidth]{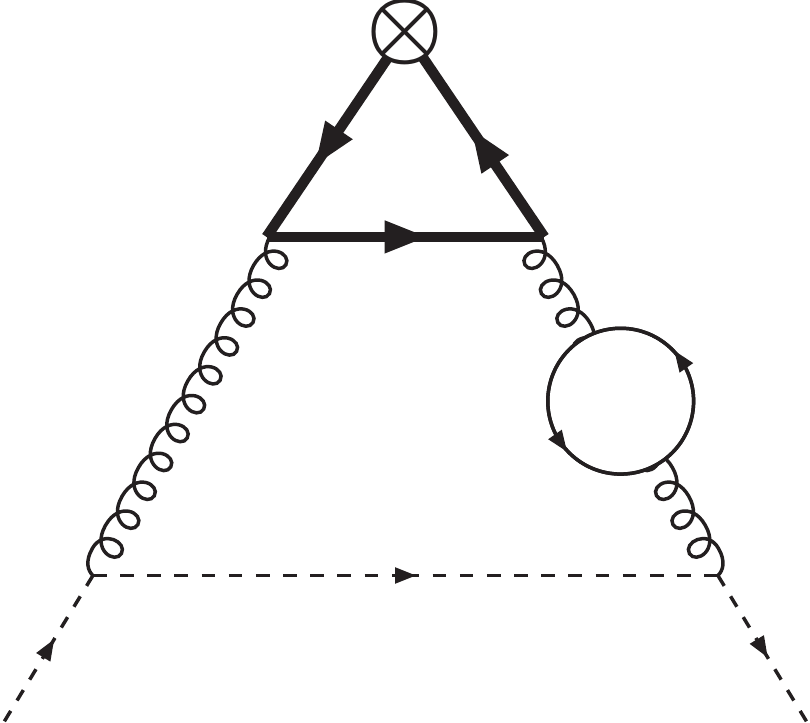}
\vspace*{-11mm}
\begin{center}
{\footnotesize (15)}
\end{center}
\end{minipage}
\hspace*{1mm}
\begin{minipage}[c]{0.19\linewidth}
     \includegraphics[width=1\textwidth]{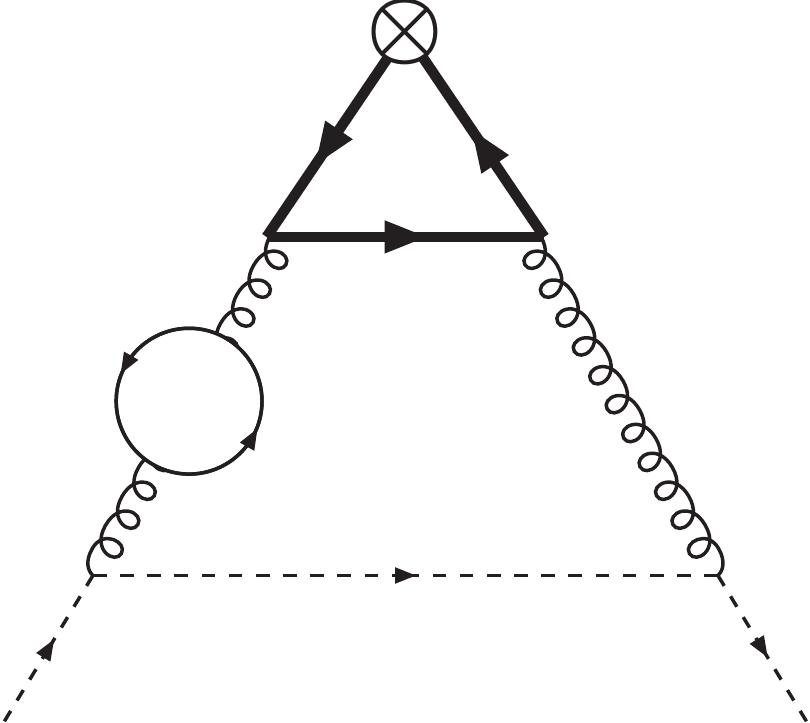}
\vspace*{-11mm}
\begin{center}
{\footnotesize (16)}
\end{center}
\end{minipage}
\caption{\sf \small Diagrams for the two-mass contributions to $\tilde{A}_{Qq}^{(3), \rm PS}$. The dashed arrow line represents the external
massless quarks, while the thick solid arrow line represents a quark of mass $m_1$, and the thin arrow line a quark of mass $m_2$. We assume $m_1 > m_2$.}
\label{diagrams}
\end{center}
\end{figure}

\vspace*{1mm}
\noindent
There are sixteen irreducible diagrams contributing to $\tilde{A}_{Qq}^{(3), {\rm PS}}$, which are shown in Figure \ref{diagrams}. 
The unrenormalized operator matrix element is obtained by adding all the diagrams and applying the quarkonic projector $P_q$ to the 
corresponding Green function $\hat{G}^{ij}_Q$,

\begin{equation}
\hat{\hat{\tilde{A}}}_{Qq}^{(3)}\left(\frac{m_1^2}{\mu^2},\frac{m_2^2}{\mu^2},\ep,N\right) = 
P_q \hat{G}^{ij}_Q \equiv
\frac{\delta^{ij}}{4 N_c} \left(\Delta.p\right)^{-N} {\rm Tr}\left[\slashed{p} \, \hat{G}^{ij}_Q\right],
\label{projector}
\end{equation}
where $p$ is the momentum of the on-shell external massless quark ($p^2=0$), 
$\Delta$ is a light-like $D$-vector, with $D = 4+ \varepsilon$, $i$ and $j$ are the color indices of each external 
leg, and $N_c$ is the number 
of colors. The diagrams, 
$D_1, \ldots, D_{16}$, are calculated directly within dimensional regularization in $D$ dimensions. These diagrams 
have the following structure,
\begin{eqnarray}
D_i(m_1,m_2,N) &=& \left(\frac{m^2}{\mu^2}\right)^{\frac{3}{2} \ep} \tilde{D}_i(\eta,\ep,N) \label{DiN} \\
&=& \left(\frac{m^2}{\mu^2}\right)^{\frac{3}{2} \ep} \int_0^1 dx \, x^{N-1} \hat{D}_i(\eta,\ep,x), \label{Dix}
\end{eqnarray}
where $N$ is the Mellin variable appearing in the Feynman rules for the operator insertions, 
cf.~\cite{Bierenbaum:2009mv,Ablinger:2017err}, $m$ is the mass of the 
heavy quark where the operator insertion is not sitting, and
\begin{equation}
\eta = \frac{m_2^2}{m_1^2},
\end{equation}
with $m_2 < m_1$, i.e.~$\eta < 1$.

In previous publications where single-mass OMEs have been computed 
\cite{Ablinger:2014lka,Behring:2014eya,Ablinger:2014uka,Ablinger:2014vwa,
Ablinger:2014nga,Behring:2015zaa,Behring:2015roa,Ablinger:2015tua,Behring:2016hpa}, 
and in our recently published two-mass calculations \cite{Ablinger:2017err},
we have always given the results both in $N$- and $x$-space. In the case of the two-mass pure singlet OME, 
finding a general $N$-space result at three loops, turns out to be rather
cumbersome. We will, therefore, present our result only in $x$-space, see Eq. (\ref{Dix}), which is anyway 
all we need in order to obtain the corresponding contribution to
the structure function $F_2(x,Q^2)$ for large values of $Q^2$, as well as the contribution to the variable 
flavor number scheme. In most of the applications one finally works in $x$-space. 

The diagrams on the first row of Figure \ref{diagrams} all give the same result, 
and the diagrams on the second row are related to the diagrams on the first row by the exchange 
$m_1 \leftrightarrow m_2$, i.e.,
\begin{eqnarray}
D_i(m_1,m_2,N) &=& D_1(m_1,m_2,N) \quad {\rm for} \quad i=2,3,4. 
\label{D2relD1} \\
D_i(m_1,m_2,N) &=& D_1(m_2,m_1,N) \quad {\rm for} \quad i=5,6,7,8. 
\label{D5relD1}
\end{eqnarray}
All of these diagrams vanish for odd values of $N$.

A relation similar to (\ref{D5relD1}) holds between the diagrams on the third row of Figure \ref{diagrams}
 and those on the fourth row. 
Furthermore, the diagrams on these two last rows which differ only in the direction of a fermion arrow are related by 
a factor of $(-1)^N$ 
relative to each other. Specifically one has
\begin{eqnarray}
D_{10}(m_1,m_2,N) &=& D_9(m_1,m_2,N). \label{D10relD9} \\
D_i(m_1,m_2,N) &=& (-1)^N D_9(m_1,m_2,N) \quad {\rm for} \quad i=11,12. \label{D11relD9} \\
D_i(m_1,m_2,N) &=& D_9(m_2,m_1,N) \quad {\rm for} \quad i=13,14. \label{D13relD9} \\
D_i(m_1,m_2,N) &=& (-1)^N D_9(m_2,m_1,N) \quad {\rm for} \quad i=15,16. \label{D15relD9}
\end{eqnarray}
We can therefore write the whole unrenormalized pure singlet operator matrix element solely in terms of diagrams 1 and 9:
\begin{eqnarray}
A_{Qq}^{(3), {\rm PS}}(N) &=& 
4 D_1(m_1,m_2,N)
+2 \left(1+(-1)^N\right) D_9(m_1,m_2,N) \nonumber \\ &&
\hspace*{-3mm}+4 D_1(m_2,m_1,N)
+2 \left(1+(-1)^N\right) D_9(m_2,m_1,N).
\label{eq:D9}
\end{eqnarray}
In Eq.~(\ref{eq:D9}) the use of the projection (\ref{projector}) is implicit.

\begin{figure}[ht]
\begin{center}
\begin{minipage}[c]{0.21\linewidth}
     \includegraphics[width=1\textwidth]{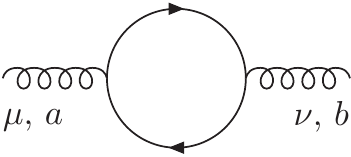}
\vspace*{-7mm}
\begin{center}
{\footnotesize ($a_1$)}
\end{center}
\end{minipage}
\hspace*{11mm}
\begin{minipage}[c]{0.21\linewidth}
     \includegraphics[width=1\textwidth]{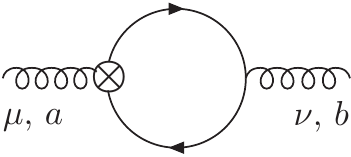}
\vspace*{-7mm}
\begin{center}
{\footnotesize ($b_1$)}
\end{center}
\end{minipage}
\hspace*{11mm}
\begin{minipage}[c]{0.21\linewidth}
     \includegraphics[width=1\textwidth]{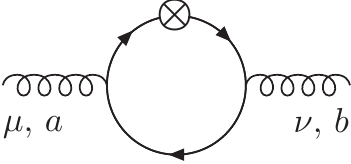}
\vspace*{-7mm}
\begin{center}
{\footnotesize ($b_2$)}
\end{center}
\end{minipage}
\caption{\sf \small Massive bubbles appearing in the Feynman diagrams shown in Figure \ref{diagrams}.}
\label{bubbles}
\end{center}
\end{figure}

All of the diagrams contain a massive fermion loop with an operator insertion (Figures \ref{bubbles}$b_1$ and \ref{bubbles}$b_2$)
and a massive bubble without the operator (Figure \ref{bubbles}$a_1$).
The latter can be rendered effectively massless by using a Mellin--Barnes
integral~\cite{MB1a,MB1b,MB2,MB3,MB4}
\begin{eqnarray}
I^{\mu\nu,ab}_{a_1}(k) &=&
g_s^2 T_F \frac{4}{\pi} \left(4 \pi\right)^{-\ep/2} \left(k_{\mu}k_{\nu}-k^2 g_{\mu\nu} \right) 
\nonumber \\ &&
\times \int_{-i\,\infty}^{+i\,\infty} d \sigma
\left(\frac{m^2}{\mu^2}\right)^{\sigma} \left(-k^2\right)^{\ep/2 - \sigma}
\frac{\Gamma(\sigma-\ep/2) \Gamma^2(2-\sigma+\ep/2) \Gamma(-\sigma)}{\Gamma(4-2 \sigma+\ep)},
\end{eqnarray}
where $\mu$ and $\nu$ are the respective Lorentz indices of the external legs, $a$ and $b$ the color indices, 
$k$ is the external momentum, $m$ is the mass of the fermion, which can be either $m_1$ or 
$m_2$, $g_s = \sqrt{4 \pi \alpha_s}$ is the strong coupling constant, and $T_F=1/2$ in $SU(N_c)$, with $N_c$ the 
number 
of colors.

Diagram 1 can then be calculated using the following expression for the massive fermion bubble containing the vertex 
operator insertion (Figure \ref{bubbles}$b_1$) :
\begin{eqnarray}
I^{\mu\nu,ab}_{b_1}(k) &=&
16 \delta_{ab} T_F g_s^2 \frac{(\Delta.k)^{N-2}}{(4\pi)^{D/2}} \Gamma(2-D/2) 
\int_0^1 dx \,\, x^N (1-x) \frac{(\Delta.k) 
\Delta_{\mu} k_{\nu} - k^2 \Delta_{\mu} \Delta_{\nu}}{(m^2-x (1-x) k^2)^{2-D/2}}.
\nonumber\\ 
\end{eqnarray}
Likewise, diagram 9 can be calculated using the following expression for the bubble containing 
the operator insertion on the fermion line (Figure \ref{bubbles}$b_2$):
\begin{eqnarray}
I^{\mu\nu,ab}_{b_2}(k) &=&
4 \delta_{ab} T_F g_s^2 \frac{(\Delta.k)^{N-2}}{(4\pi)^{D/2}} \int_0^1 dx \,\, x^{N-2} (1-x) \biggl[ 
\nonumber \\ &&
-2\left(x (1-x) (g_{\mu \nu} k^2 -2 k_{\mu} k_{\nu})+ m^2 g_{\mu \nu}\right) \frac{x^2 \Gamma(3-D/2) 
(\Delta.k)^2}{(m^2-x (1-x) k^2)^{3-D/2}} 
\nonumber \\ &&
+\Gamma(2-D/2) (2 N x+1-N) \frac{x (k_{\mu} \Delta_{\nu}+k_{\nu} \Delta_{\mu})(\Delta.k)}{(m^2-x (1-x) k^2)^{2-D/2}} 
\nonumber \\ &&
+\Gamma(2-D/2) ((N-1) (1-2 x)-D x) \frac{x g_{\mu \nu} (\Delta.k)^2}{(m^2-x (1-x) k^2)^{2-D/2}} 
\nonumber \\ &&
-\Gamma(1-D/2) \frac{N-1}{1-x} (N (1-x)-1) \frac{\Delta_{\mu} \Delta_{\nu}}{(m^2-x (1-x) k^2)^{1-D/2}} \biggr].
\end{eqnarray}
After inserting these expressions, we introduced the corresponding projector for a quarkonic OME given in Eq. 
(\ref{projector}), and performed the Dirac matrix algebra and the trace
arising in the numerator of the diagrams using the program {\tt FORM} \cite{FORM}. This leads to a linear combination 
of integrals, the denominators of which were then combined
using Feynman parameters. The momentum integrals were then performed, and we obtained expressions 
where one of the Feynman parameter integrals is
already in the form of a Mellin transform. Therefore, we left this Feynman parameter un-integrated, and integrated the remaining ones, after which we obtained the following 
expression for diagram 1,

\begin{equation}
D_1(m_1,m_2,N) = -128 C_F T_F^2 \left(1+(-1)^N\right) \big(J_1-J_2\big),
\end{equation}
with $C_F = (N_c^2-1)/(2 N_c)$ and $C_A = N_c$, and
\begin{eqnarray}
J_1 &=& 
\left(\frac{m_1^2}{\mu^2}\right)^{\frac{3}{2} \varepsilon} \frac{\Gamma (N-1)}{\Gamma \left(N+\frac{\varepsilon 
}{2}\right)} 
\int_0^1 dx \, x^{N+\frac{\varepsilon}{2}} (1-x)^{1+\frac{\varepsilon}{2}} B_1\left(\frac{\eta}{x (1-x)}\right),
\label{J1} \\ 
J_2 &=& 
\left(\frac{m_1^2}{\mu^2}\right)^{\frac{3}{2} \varepsilon} \frac{\Gamma (N)}{\Gamma \left(N+1+\frac{\varepsilon 
}{2}\right)} 
\int_0^1 dx \, x^{N+\frac{\varepsilon}{2}} (1-x)^{1+\frac{\varepsilon}{2}} B_1\left(\frac{\eta}{x (1-x)}\right).
\label{J2}
\end{eqnarray}
Here $B_1(\xi)$ is the following contour integral,
\begin{equation}
B_1(\xi) = \frac{1}{2 \pi i} \int_{-i \infty}^{i \infty} d\sigma \, \xi^{\sigma} \,
\Gamma(-\sigma) \Gamma(-\sigma+\varepsilon) \Gamma\left(\sigma-\frac{3 \varepsilon}{2}\right) \Gamma\left(\sigma -\frac{\varepsilon}{2}\right)
\frac{\Gamma^2(\sigma+2-\varepsilon)}{\Gamma(2 \sigma+4-2 \varepsilon)}.
\label{B1}
\end{equation}
For diagram 9, we get
\begin{eqnarray}
D_9(m_1,m_2,N) &=& 64 C_F T_F^2 \biggl\{
-\frac{1}{4} (\ep+2) \big[-2 J_3+2 \eta J_4+(2 N+2+\ep) J_5
\nonumber \\ &&
-(N-1) J_6\big]
+N (N-1) \big(J_7-J_8\big)
-(N-1) \big(J_9-J_{10}\big)
\biggr\},
\label{D9}
\end{eqnarray}
where
\begin{eqnarray}
J_3 &=& 
\left(\frac{m_1^2}{\mu^2}\right)^{\frac{3}{2} \varepsilon} 
\frac{\Gamma(N+1)}{\Gamma\left(N+2+\frac{\varepsilon}{2}\right)} 
\int_0^1 dx \, x^{N+\frac{\varepsilon}{2}} (1-x)^{1+\frac{\varepsilon}{2}} B_2\left(\frac{\eta}{x (1-x)}\right), 
\label{J3} \\ 
J_4 &=& 
\left(\frac{m_1^2}{\mu^2}\right)^{\frac{3}{2} \varepsilon} 
\frac{\Gamma(N+1)}{\Gamma\left(N+2+\frac{\varepsilon}{2}\right)} 
\int_0^1 dx \, x^{N-1+\frac{\varepsilon}{2}} (1-x)^{\frac{\varepsilon}{2}} B_3\left(\frac{\eta}{x (1-x)}\right),
\label{J4} \\
J_5 &=& 
\left(\frac{m_1^2}{\mu^2}\right)^{\frac{3}{2} \varepsilon} 
\frac{\Gamma(N+1)}{\Gamma\left(N+2+\frac{\varepsilon}{2}\right)} 
\int_0^1 dx \, x^{N+\frac{\varepsilon}{2}} (1-x)^{1+\frac{\varepsilon}{2}} B_1\left(\frac{\eta}{x (1-x)}\right),
\label{J5} \\ 
J_6 &=& 
\left(\frac{m_1^2}{\mu^2}\right)^{\frac{3}{2} \varepsilon} 
\frac{\Gamma(N+1)}{\Gamma\left(N+2+\frac{\varepsilon}{2}\right)} 
\int_0^1 dx \, x^{N-1+\frac{\varepsilon}{2}} (1-x)^{1+\frac{\varepsilon}{2}} B_1\left(\frac{\eta}{x (1-x)}\right),
\label{J6} \\ 
J_7 &=& 
\left(\frac{m_1^2}{\mu^2}\right)^{\frac{3}{2} \varepsilon} 
\frac{\Gamma(N-1)}{\Gamma\left(N+\frac{\varepsilon}{2}\right)} 
\int_0^1 dx \, x^{N-1+\frac{\varepsilon}{2}} (1-x)^{2+\frac{\varepsilon}{2}} B_4\left(\frac{\eta}{x (1-x)}\right),
\label{J7} \\
J_8 &=& 
\left(\frac{m_1^2}{\mu^2}\right)^{\frac{3}{2} \varepsilon} 
\frac{\Gamma(N)}{\Gamma\left(N+1+\frac{\varepsilon}{2}\right)} 
\int_0^1 dx \, x^{N-1+\frac{\varepsilon}{2}} (1-x)^{2+\frac{\varepsilon}{2}} B_4\left(\frac{\eta}{x (1-x)}\right),
\label{J8} \\
J_9 &=& 
\left(\frac{m_1^2}{\mu^2}\right)^{\frac{3}{2} \varepsilon} 
\frac{\Gamma(N-1)}{\Gamma\left(N+\frac{\varepsilon}{2}\right)} 
\int_0^1 dx \, x^{N-1+\frac{\varepsilon}{2}} (1-x)^{1+\frac{\varepsilon}{2}} B_4\left(\frac{\eta}{x (1-x)}\right),
\label{J9} \\
J_{10} &=& 
\left(\frac{m_1^2}{\mu^2}\right)^{\frac{3}{2} \varepsilon} 
\frac{\Gamma(N)}{\Gamma\left(N+1+\frac{\varepsilon}{2}\right)} 
\int_0^1 dx \, x^{N-1+\frac{\varepsilon}{2}} (1-x)^{1+\frac{\varepsilon}{2}} B_4\left(\frac{\eta}{x (1-x)}\right), 
\label{J10} 
\end{eqnarray}
with
\begin{eqnarray} 
B_2(\xi) &=& \frac{1}{2 \pi i} \int_{-i \infty}^{i \infty} d\sigma \, \xi^{\sigma} \,
\Gamma(-\sigma) \Gamma(-\sigma+\varepsilon) \Gamma\left(\sigma-\frac{3 \varepsilon}{2}\right) \Gamma\left(\sigma+1-\frac{\varepsilon}{2}\right)
\frac{\Gamma^2(\sigma+2-\varepsilon)}{\Gamma(2 \sigma+4-2 \varepsilon)},
\nonumber \\
\label{B2}
\\ 
B_3(\xi) &=& \frac{1}{2 \pi i} \int_{-i \infty}^{i \infty} d\sigma \, \xi^{\sigma} \,
\Gamma(-\sigma) \Gamma(-\sigma-1+\varepsilon) \Gamma\left(\sigma+1-\frac{3 \varepsilon}{2}\right) \Gamma\left(\sigma+1-\frac{\varepsilon}{2}\right)
\nonumber \\ && \phantom{\frac{1}{2 \pi i} \int_{-i \infty}^{i \infty} d\sigma} \times
\frac{\Gamma^2(\sigma+3-\varepsilon)}{\Gamma(2 \sigma+6-2 \varepsilon)},
\label{B3}
\\
B_4(\xi) &=& \frac{1}{2 \pi i} \int_{-i \infty}^{i \infty} d\sigma \, \xi^{\sigma} \,
\Gamma(-\sigma) \Gamma(-\sigma+\varepsilon) \Gamma\left(\sigma-\frac{3 \varepsilon}{2}\right) \Gamma\left(\sigma-1-\frac{\varepsilon}{2}\right)
\frac{\Gamma^2(\sigma+2-\varepsilon)}{\Gamma(2 \sigma+4-2 \varepsilon)}.
\nonumber \\
\label{B4}
\end{eqnarray}

We want to compute the diagrams, and therefore the integrals (\ref{B1}, \ref{B2}--\ref{B4}), as an expansion in $\ep$ 
up to $O(\ep^0)$. 
Notice that there is always a factor consisting on a ratio of $\Gamma$--functions depending on $N$ in these 
integrals, and there are also additional factors
of $N$ in Eq. (\ref{D9}). After the aforementioned $\ep$ expansion is performed, some of the integrals will be left with a factor of the form
\begin{equation}
\frac{1}{N+l}, \quad {\rm with} \quad l \in \{-1,0,1\}.
\end{equation} 
In order to get our results really in terms of a Mellin transform, these factors need to be absorbed into the integral in $x$, 
which can be accomplished using integration by parts,
\begin{eqnarray}
\frac{1}{N+l} \int_a^b dx \, x^{N-1} f(x) &=& 
\frac{b^{N+l}}{N+l} \int_a^b dy \frac{f(y)}{y^{l+1}} - \int_a^b dx \, x^{N+l-1} \int_a^x dy \frac{f(y)}{y^{l+1}} 
\label{absorbN1}
\\ &=&
\frac{a^{N+l}}{N+l} \int_a^b dy \frac{f(y)}{y^{l+1}} + \int_a^b dx \, x^{N+l-1} \int_x^b dy \frac{f(y)}{y^{l+1}}.
\label{absorbN2}
\end{eqnarray}
We will see later how this is manifested in the final result.
\subsection{Computation of the contour integrals}
\label{sec:32}

\vspace*{1mm}
\noindent
We have seen that the diagrams where the operator insertion lies on the heaviest quark are related to the ones where the
insertion lies on the lightest internal quark by the exchange $m_1 \leftrightarrow m_2$, which means that in order to 
go from the
former diagrams to the latter, we also need to do the change $\eta \rightarrow 1/\eta$. Therefore, while in the case of
diagrams $D_1,\ldots,D_4$ and $D_9,\ldots,D_{12}$, we need to evaluate the contour integrals (\ref{B1}, \ref{B2}--\ref{B4}) with
\begin{equation}
\, \xi = \frac{\eta}{x (1-x)},
\label{xidef1}
\end{equation}
in the case of diagrams $D_5,\ldots,D_8$ and $D_{13},\ldots,D_{16}$, we have instead
\begin{equation}
\xi = \frac{1}{\eta x (1-x)}.
\label{xidef2}
\end{equation}
As we will see, the contour integrals arising in the case of Eq. (\ref{xidef1}) will require a different treatment to those in the case of Eq. (\ref{xidef2}).

The integrals (\ref{B1}, \ref{B2}--\ref{B4}) can be computed with the help of the Mathematica package {\tt MB} \cite{MB}, 
together with the add-on package {\tt MBresolve} \cite{MBr}. 
These packages allow us to resolve the singularity structure in $\ep$ of these integrals by taking residues in $\sigma$, after which we can 
perform the expansion in $\ep$. This leads to expressions of the form
\begin{equation}
B_i(\xi) = B_i^{(\ep)}(\xi)+B_i^{(0)}(\xi), \quad i=1,2,3,4,
\label{epreg}
\end{equation}
where the functions $B_i^{(\ep)}(\xi)$ are the sum of residues. For example, in the case of $B_1(\xi)$, we must take 
residues at $\sigma = \ep/2$ and $\sigma = 3 \ep/2$, which leads to

\begin{eqnarray}
B_1^{(\ep)}(\xi) &=& \xi^{\ep/2} \Gamma (-\ep) \Gamma \left(-\frac{\ep}{2}\right) \Gamma \left(\frac{\ep}{2}\right) 
\frac{\Gamma^2\left(2-\frac{\ep}{2}\right)}{\Gamma (4-\ep)} \nonumber \\ &&
+\xi^{3 \ep/2} \Gamma \left(-\frac{3 \ep}{2}\right) \Gamma \left(-\frac{\ep}{2}\right) \Gamma (\ep) 
\frac{\Gamma^2\left(2+\frac{\ep}{2}\right)}{\Gamma (4+\ep)}.
\end{eqnarray}

Once the residues are taken, it is possible to expand the original integrals in $\ep$. This leads to the contour integrals that we denote by $B_i^{(0)}(\xi)$,
for which the {\tt MBresolve} package finds a contour consisting of a line parallel to the imaginary axis with a 
real value for $\sigma \in [-1,0]$\footnote{Here we have chosen this value to be $-1/2$,
but any value will do, since the poles of the integrands are at integer values of $\sigma$.}. The functions are given 
by
\begin{eqnarray}
B_1^{(0)}(\xi) &=& \frac{1}{2 \pi i} \int_{-1/2-i \infty}^{-1/2+i \infty} d\sigma \, \xi^{\sigma} \,
\Gamma^2(-\sigma) \Gamma^2(\sigma)
\frac{\Gamma^2(\sigma+2)}{\Gamma(2 \sigma+4)}+O(\ep),
\label{B10} \\ 
B_2^{(0)}(\xi) &=& \frac{1}{2 \pi i} \int_{-1/2-i \infty}^{-1/2+i \infty} d\sigma \, \xi^{\sigma} \,
\Gamma^2(-\sigma) \Gamma(\sigma) \Gamma(\sigma+1)
\frac{\Gamma^2(\sigma+2)}{\Gamma(2 \sigma+4)}+O(\ep),
\label{B20}\\ 
B_3^{(0)}(\xi) &=& \frac{1}{2 \pi i} \int_{-1/2-i \infty}^{-1/2+i \infty} d\sigma \, \xi^{\sigma} \,
\Gamma(-\sigma) \Gamma(-\sigma-1) \Gamma^2(\sigma+1)
\frac{\Gamma^2(\sigma+3)}{\Gamma(2 \sigma+6)}+O(\ep),
\label{B30}\\ 
B_4^{(0)}(\xi) &=& \frac{1}{2 \pi i} \int_{-1/2-i \infty}^{-1/2+i \infty} d\sigma \, \xi^{\sigma} \,
\Gamma^2(-\sigma) \Gamma(\sigma) \Gamma(\sigma-1)
\frac{\Gamma^2(\sigma+2)}{\Gamma(2 \sigma+4)}+O(\ep).
\label{B40}
\end{eqnarray}

In order to calculate the contour integrals (\ref{B10}--\ref{B40}), we need to close the contour either to the right 
or to the left, 
depending on the convergence of the resulting sum of residues. The integrand always contains a $\Gamma$-function in 
the
denominator that can be rewritten using Legendre's  duplication formula,
\begin{equation}
\Gamma(2 \sigma+2 l) = \frac{4^{\sigma+l}}{2 \sqrt{\pi}} \Gamma(\sigma+l) \Gamma\left(\sigma+l+\frac{1}{2}\right), \quad l=2,3,
\end{equation}
which means that if we close the contour to the right, 
the sum of residues will be convergent if and only if $\xi<4$, while if we close the contour to the left, the sum of residues will
be convergent if and only if $\xi>4$.
Therefore, in the case where $\xi = (\eta x (1-x))^{-1}$, we just need to close the contour to the left, since 
$\eta<1$, and therefore $\xi \geq 4/\eta > 4$.
The calculation in the case where $\xi = \eta/(x (1-x))$, on the other hand, is a bit more complicated, since in this case we need to split the integration range in $x$ into 
the different regions where $\xi$ can be bigger or smaller than 4,
\begin{eqnarray}
\frac{\eta}{x (1-x)} &<& 4, \quad 
{\rm for} \;\;\; x \in \left(\eta_-,\eta_+\right), \\
\frac{\eta}{x (1-x)} &>& 4, \quad 
{\rm for} \;\;\; x \in \left(0,\eta_-\right) \quad {\rm or} \;\;\;\; x \in \left(\eta_+,1\right),
\end{eqnarray}
where
\begin{equation}
\eta_{\pm} = \frac{1}{2}\left(1 \pm \sqrt{1-\eta}\right).
\end{equation}
Therefore, in this case, we will need to close the contour to the right when $\eta_-<x<\eta_+$, and to the left when $0<x<\eta_-$ or $\eta_+<x<1$.

Closing the contour to the right and summing residues, we obtain
\begin{eqnarray}
B_1^{(0)}(\xi) &=& -\left(\frac{\zeta_2}{6}+\frac{14}{27}\right) \ln(\xi)
-\frac{1}{36} \ln^3(\xi)
+\frac{5}{36} \ln^2(\xi)
+\frac{5}{18} \zeta_2
-\frac{\zeta_3}{3}
+\frac{82}{81} 
\nonumber \\ &&
+\sum_{k=1}^{\infty} \xi^k \frac{\Gamma^2(k+2)}{k^2 \Gamma (2 k+4)} \left(
2 S_1(2 k+3)
-2 S_1(k)
+\frac{2}{k (k+1)}
-\ln(\xi)\right),
\\
B_2^{(0)}(\xi) &=& -\frac{1}{12} \ln^2(\xi)
+\frac{5}{18} \ln(\xi)
-\frac{\zeta_2}{6}
-\frac{14}{27}
+\sum_{k=1}^{\infty} \xi^k \frac{\Gamma^2(k+2)}{k \Gamma (2 k+4)} \biggl(
2 S_1(2 k+3)
\nonumber \\ &&
-2 S_1(k) 
-\frac{k-1}{k (k+1)}
-\ln(\xi)
\biggr), 
\\
B_3^{(0)}(\xi) &=& \sum_{k=0}^{\infty} \xi^k \frac{\Gamma^2(k+3)}{(k+1) \Gamma (2 k+6)} \biggl(
2 S_1(k+2)
-2 S_1(2 k+5)
-\frac{1}{k+1}
+\ln(\xi)
\biggr),
\\
B_4^{(0)}(\xi) &=& -\left(\frac{\xi}{30}+\frac{1}{9}\right) \zeta_2
+\left(\frac{107 \xi}{900}+\frac{\zeta_2}{6}+\frac{11}{27}\right) \ln(\xi)
-\frac{4067}{13500} \xi 
+\frac{1}{36} \ln^3(\xi) 
\nonumber \\ &&
-\left(\frac{\xi}{60}+\frac{1}{18}\right) \ln^2(\xi)
+\frac{\zeta_3}{3}
-\frac{49}{81}
+\sum_{k=2}^{\infty} \xi^k \frac{\Gamma^2(k+2)}{(k-1) k^2 \Gamma (2 k+4)} \biggl(
2 S_1(2 k+3) 
\nonumber \\ &&
-2 S_1(k) 
+\frac{k^2+3 k-2}{(k-1) k (k+1)}
-\ln(\xi)
\biggr),
\end{eqnarray}
while closing to the left leads to
\begin{eqnarray}
B_1^{(0)}(\xi) &=&  \frac{1}{\xi}\ln(\xi)
+\sum_{k=2}^{\infty} \xi^{-k} \frac{\Gamma (2 k-3)}{k^2 \Gamma^2(k-1)} \biggl[
\ln(\xi) \left(4 S_1(k)-4 S_1(2 k-4)-\frac{4}{k-1}\right)  \nonumber \\ &&
+\ln^2(\xi)
+4 S_1^2(k)+4 S_1^2(2 k-4)
-8 S_1(k) \left(S_1(2 k-4)+\frac{1}{k-1}\right) \nonumber \\ &&
+\frac{8}{k-1} S_1(2 k-4)
+2 S_2(k)
-4 S_2(2 k-4)
+\frac{2}{(k-1)^2}
+2 \zeta_2
\biggr],
\\
B_2^{(0)}(\xi) &=& \frac{1}{\xi}-\frac{\ln(\xi)}{\xi}
+\sum_{k=2}^{\infty} \xi^{-k} \frac{\Gamma (2 k-3)}{k \Gamma^2(k-1)}  \biggl[
\ln(\xi) \biggl(4 S_1(2 k-4)-4 S_1(k) 
+\frac{2 (3 k-1)}{(k-1) k}\biggr) \nonumber \\ &&
-\ln^2(\xi)
-2 S_2(k)
-4 S_1^2(k)
-4 S_1^2(2 k-4)
+4 S_2(2 k-4) 
-\frac{2 (3 k-2)}{(k-1)^2 k} \nonumber \\ &&
-\frac{4 (3 k-1)}{(k-1) k} S_1(2 k-4)
+4 S_1(k) \left(2 S_1(2 k-4)+\frac{3 k-1}{(k-1) k}\right)
-2 \zeta_2
\biggr], 
\\
B_3^{(0)}(\xi) &=& -\frac{14}{27 \xi}
-\frac{\zeta_2}{6 \xi}
+\frac{5}{18 \xi} \ln(\xi)
-\frac{1}{12 \xi} \ln^2(\xi)
-\frac{1}{\xi^2}
+\frac{1}{\xi^2} \ln(\xi) \nonumber  \\ &&
+\sum_{k=3}^{\infty} \xi^{-k} \frac{(k-2) \Gamma (2 k-5)}{\Gamma (k-2) \Gamma (k)} \biggl[
\ln(\xi) \left(4 S_1(k-3)-4 S_1(2 k-6)+\frac{2}{k-1}\right) \nonumber  \\ &&
+4 S_1^2(k-3)
+4 S_1^2(2 k-6)
+4 S_1(k-3) \left(\frac{1}{k-1}-2 S_1(2 k-6)\right) 
+2 \zeta_2 \nonumber  \\ &&
-\frac{4}{k-1} S_1(2 k-6)
+2 S_2(k-3)
-4 S_2(2 k-6)
+\frac{2}{(k-1)^2}
+\ln^2(\xi)
\biggr], 
\\
B_4^{(0)}(\xi) &=& -\frac{1}{4 \xi}-\frac{\ln(\xi)}{2 \xi}
+\sum_{k=2}^{\infty} \xi^{-k} \frac{(k-1) \Gamma (2 k-3)}{k \Gamma (k-1) \Gamma (k+2)} \biggl[ 
4 S_2(2 k-4)
-S_2(k+1) \nonumber \\ &&
-\ln^2(\xi)
+\ln(\xi) \biggl(4 S_1(2 k-4)
-2 S_1(k)
-2 S_1(k+1)
+\frac{4}{k-1}\biggr) \nonumber \\ &&
-4 S_1^2(2 k-4) 
-\frac{8}{k-1} S_1(2 k-4)
+4 S_1(k+1) \left(S_1(2 k-4)+\frac{1}{k-1}\right) 
\nonumber \\ &&
+S_1(k) \left(4 S_1(2 k-4)-2 S_1(k+1)+\frac{4}{k-1}\right)
-S_1^2(k)
-S_1^2(k+1)
\nonumber \\ &&
-S_2(k)
-\frac{2}{(k-1)^2}
-2 \zeta_2
\biggr].
\end{eqnarray}
Here $S_{\vec{a}} \equiv S_{\vec{a}}(N)$ denotes the (nested) harmonic sums \cite{HSUM}
\begin{eqnarray}
S_{b,\vec{a}}(N) = \sum_{k=1}^N \frac{({\rm sign}(b))^k}{k^{|b|}} S_{\vec{a}}(k),~~~S_\emptyset = 1,~~~b,a_i \in \mathbb{Z} 
\backslash \{0\}~.
\end{eqnarray}
In the above expressions ratios of $\Gamma$-functions are related to special binomial coefficients, like
\begin{eqnarray}
\frac{\Gamma^2(k+1)}{\Gamma(2k+2)} = \frac{1}{2k} \frac{1}{\displaystyle \binom{2k}{k}}.
\end{eqnarray}
All of the above sums can be performed using the Mathematica packages {\tt Sigma}~\cite{SIG1,SIG2}, {\tt 
HarmonicSums}~\cite{HARMONICSUMS,Ablinger:2011te,Ablinger:2013cf},
{\tt EvaluateMultiSums} and {\tt SumProduction} \cite{EMSSP}.
The results are expressed in terms of generalized iterated integrals.

\begin{equation}
G\left(\left\{f_1(\tau),f_2(\tau),\cdots,f_n(\tau)\right\},z\right)
=\int_0^z  d\tau_1~f_1(\tau_1)  
G\left(\left\{f_2(\tau),\cdots,f_n(\tau)\right\},\tau_1\right),
\label{Gfunctions}
\end{equation}
with
\begin{equation}
 G\Biggl(\Biggl\{\underbrace{\frac{1}{\tau},\frac{1}{\tau},
  \cdots,\frac{1}{\tau}}_{\text{n times}}\Biggr\},z\Biggr)
\equiv
\frac{1}{n!} \ln^n(z)~.
\end{equation}
In principle, the letters in the alphabet of these iterated integrals (i.e., the functions $f_k(\tau)$) can be 
any function (or distribution), for which the iterated integral exists. 
In the particular case where the letters are restricted to $\frac{1}{\tau}$, $\frac{1}{1-\tau}$ and $\frac{1}{1+\tau}$, these
integrals correspond to the harmonic polylogarithms \cite{Remiddi:1999ew}, which are defined by
\begin{eqnarray}
H_{b,\vec{a}}(x) &=& \int_0^x dy f_b(y) H_{\vec{a}}(y),~~~H_\emptyset = 1,~a_i, b~\in~\{0,1,-1\}~,
\end{eqnarray}
with
\begin{eqnarray}
f_0(x) = \frac{1}{x},~~~f_1(x) = \frac{1}{1-x},~~~f_{-1}(x) = \frac{1}{1+x}~,
\end{eqnarray}
and
\begin{equation}
H_{\underbrace{\text{\scriptsize 0},\ldots,\text{\scriptsize 0}}_{\text{\scriptsize n times}}}(x)
=  G\Biggl(\Biggl\{\underbrace{\frac{1}{\tau},\frac{1}{\tau},
  \cdots,\frac{1}{\tau}}_{\text{n times}}\Biggr\},x\Biggr)
=
\frac{1}{n!} \ln^n(x)~.
\end{equation}
We will see that not only harmonic polylogarithms appear in our final result, but also iterated integrals with square 
roots in the letters.\footnote{Root-valued iterated integrals have been 
discussed before in Ref.~\cite{Ablinger:2014bra}.}

\section{The massive operator matrix element}
\label{sec:4}

\vspace*{1mm}
\noindent
We obtain the following expression for the $O(\ep^0)$ term of the unrenormalized 3-loop two-mass pure singlet operator 
matrix element 
\begin{eqnarray}
\tilde{a}_{Qq}^{(3), \rm PS}(x) &=& C_F T_F^2 \Biggl\{
R_0(m_1,m_2,x) +\big(\theta(\eta_--x)+\theta(x-\eta_+)\big) x \, g_0(\eta,x)
\nonumber \\ &&
+\theta(\eta_+-x) \theta(x-\eta_-) \biggl[x \, f_0(\eta,x)
\nonumber \\ &&
-\int_{\eta_-}^x dy \left(f_1(\eta,y)+\frac{y}{x} f_2(\eta,y)+\frac{x}{y} f_3(\eta,y)\right)\biggr] 
\nonumber \\ &&
+\theta(\eta_--x) \int_x^{\eta_-} dy \left(g_1(\eta,y)+\frac{y}{x} g_2(\eta,y)+\frac{x}{y} g_3(\eta,y)\right) 
\nonumber \\ &&
-\theta(x-\eta_+) \int_{\eta_+}^x dy \left(g_1(\eta,y)+\frac{y}{x} g_2(\eta,y)+\frac{x}{y} g_3(\eta,y)\right) 
\nonumber \\ &&
+x \, h_0(\eta,x) +\int_x^1 dy \left(h_1(\eta,y)+\frac{y}{x} h_2(\eta,y)+\frac{x}{y} h_3(\eta,y)\right) 
\nonumber \\ &&
+\theta(\eta_+-x) \int_{\eta_-}^{\eta_+} dy \left(f_1(\eta,y)+\frac{y}{x} f_2(\eta,y)+ \frac{x}{y} f_3(\eta,y)\right) 
\nonumber \\ &&
+\int_{\eta_+}^1 dy \left(g_1(\eta,y)+\frac{y}{x} g_2(\eta,y)+\frac{x}{y} g_3(\eta,y)\right)
\Biggr\}.
\label{aQq}
\end{eqnarray}
Here $\theta(z)$ denotes the Heaviside function
\begin{eqnarray}
\theta(z) = \left\{\begin{array}{ll} 1 &~~z \geq 0\\
                                     0 &~~z < 0.
\end{array} \right.
\end{eqnarray}
The function $R_0(m_1,m_2,x)$ arises from the residues taken in order to resolve the singularities in $\ep$ of the 
contour integrals,
see Eq. (\ref{epreg}). The functions $f_i(\eta,x)$, $g_i(\eta,x)$ and $h_i(\eta,x)$, with $i=0,1,2,3$, arise from the sum of residues
of the contour integrals that remain after the $\ep$ expansion, as described in the previous section. The functions with $i=0$ are 
those where no additional factor depending on $N$ needed to be absorbed. The functions with $i=1$, $i=2$ and $i=3$ are those where a 
factor of $1/N$, $1/(N-1)$ and $1/(N+1)$ was absorbed, respectively, see Eqs.~(\ref{absorbN1}, \ref{absorbN2}). The 
different
Heaviside $\theta$ functions restrict the corresponding values of $x$ to the appropriate regions.

Since no contour integral needs to be performed in the case $R_0(m_1,m_2,x)$, the easiest way to compute this function is to integrate
in $x$ and then perform the Mellin inversion using {\tt HarmonicSums}. We obtain,
\begin{eqnarray}
R_0(m_1,m_2,x) &=& 
32 \left(L_1^3+L_1^2 L_2+L_1 L_2^2+L_2^3\right) \Biggl[\frac{P_0}{3 x}-2 (x+1) H_0\Biggr]
\nonumber \\ &&
+32 \big(L_1^2+L_2^2\big)
\Biggl[
2 (x+1) \biggl(
H_{0,0}
+\frac{1}{3} H_{0,1}
-\frac{\zeta_2}{3}
\biggr)
-\frac{P_0 H_1}{9 x}
\nonumber \\ &&
-\frac{1}{9} (4 x+5) (7 x+5) H_0
+\frac{x-1}{27 x} \big(170 x^2+53 x+80\big)
\Biggr]
\nonumber \\ &&
+128 L_1 L_2 \Biggl[
\frac{x-1}{27 x} \big(56 x^2+47 x+20\big)
+\frac{2}{3} (x+1) \big(H_{0,1}-\zeta_2\big)
\nonumber \\ &&
-\frac{4}{9} \big(x^2+7 x+4\big) H_0
-\frac{P_0 H_1}{9 x}
\Biggr]
+\frac{128 \zeta_3}{27 x} \big(64 x^3+35 x^2-25 x+8\big)
\nonumber \\ &&
+64 (L_1+L_2) \Biggl[
(x+1) \biggl(
\frac{29}{9} H_{0,1}
-\frac{4}{3} H_{0,0,0}
+\frac{2}{3} H_{0,0,1}
+2 H_{0,1,0}
\nonumber \\ &&
-\frac{4}{3} H_{0,1,1}
-\frac{8}{3} \zeta_2 H_0
+\frac{14}{3} \zeta_3
\biggr)
+\frac{x-1}{27 x} \big(260 x^2+231 x+116\big)
\nonumber \\ &&
+\frac{P_0}{3 x} \biggl(
\frac{2}{3} H_{1,1}
-H_{1,0}
\biggr)
+\frac{2 \zeta_2}{9 x} \big(6 x^3-10 x^2-19 x-6\big)
\nonumber \\ &&
-\frac{1}{27} \big(168 x^2+265 x+229\big) H_0
-\frac{4 (x-1)}{27 x} \big(5 x^2+23 x+5\big) H_1
\nonumber \\ &&
+\frac{2}{9} \big(6 x^2+4 x-5\big) H_{0,0}
\Biggr]
+\frac{64 \zeta_2}{81 x} \big(282 x^3-229 x^2-85 x-120\big)
\nonumber \\ &&
+\frac{64 P_0}{9 x} \biggl(
4 H_{1,0,0}
-2 H_{1,0,1}
-2 H_{1,1,0}
-\frac{2}{3} H_{1,1,1}
+\zeta_2 H_1
\biggr)
\nonumber \\ &&
+128 (x+1) \biggl(
        \frac{2}{9} (6 x-5) H_{0,1,0}
        +\frac{8}{9} H_{0,0,0,0}
        -\frac{4}{9} H_{0,0,0,1}
        -\frac{4}{3} H_{0,0,1,0}
\nonumber \\ &&
        +\frac{2}{9} H_{0,0,1,1}
        -\frac{4}{3} H_{0,1,0,0}
        +\frac{2}{3} H_{0,1,0,1}
        -2 \zeta_3 H_0
        +\frac{2}{3} H_{0,1,1,0}
        +\frac{2}{9} H_{0,1,1,1}
\nonumber \\ &&
        +\frac{7}{9} \zeta_2 H_{0,0}
        -\frac{1}{3} \zeta_2 H_{0,1}
        +\frac{8}{15} \zeta_2^2
\biggr)%
-\frac{128}{27} \big(12 x^2+19 x+19\big) H_{0,1,1}
\nonumber \\ &&
-\biggl(
\frac{256}{243} \big(813 x^2+29 x+263\big)
+\frac{64}{27} \zeta_2 \big(60 x^2+91 x+37\big)
\biggr) H_0
\nonumber \\ &&
+\frac{128 (x-1)}{81 x} \big(22 x^2-25 x+4\big) H_1
+\frac{256}{81} \big(84 x^2+109 x+100\big) H_{0,0}
\nonumber \\ &&
+\frac{256}{27} \big(6 x^2-5 x-5\big) H_{0,0,1}
-\frac{128 (x-1)}{27 x} \big(56 x^2-43 x+20\big) H_{1,0}
\nonumber \\ &&
+\frac{128 (x-1)}{81 x} \big(40 x^2+49 x+40\big) H_{1,1}
-\frac{256}{27} \big(12 x^2-x-10\big) H_{0,0,0}
\nonumber \\ &&
+\frac{128}{81} (47 x+29) H_{0,1}
+\frac{256 (x-1)}{729 x} \big(2602 x^2-203 x+1360\big),
\label{eq:R00}
\end{eqnarray}
where
\begin{equation}
P_0 = (x-1) (4+7 x+4 x^2),
\end{equation}
$L_1$ and $L_2$ are the logarithms defined in Eq. (\ref{L1L2}), and we used the shorthand notation 
$H_{\vec{a}}(x) \equiv H_{\vec{a}}$. In principle (\ref{eq:R00}) could still be reduced to a shorter basis using 
shuffle-algebra \cite{Blumlein:2003gb}.

The $f_i(\eta,x)$ functions, which are defined in the range $\eta_- < x <\eta_+$, are given by
\begin{eqnarray}
f_0(\eta,x) &=&
\frac{8 P_3 \big(4 x (1-x)-\eta\big)^{3/2}}{45 \eta ^{3/2} (x-1) x^3} K_1\left(\frac{\eta }{x (1-x)}\right)
-\frac{16 (x-1)}{3 x} \Biggl\{
K_2\left(\frac{\eta }{x (1-x)}\right)
\nonumber \\ &&
-\frac{2 \left(6 \eta +30 x^2-5 x\right)}{15 x} \left[2 \zeta_2+\ln^2\left(\frac{\eta }{x (1-x)}\right)\right]
\Biggr\}
+\frac{P_1}{90 (x-1)^3 x^5}
\nonumber \\ &&
+\frac{2 P_2}{45 (x-1)^3 x^5} \ln \left(\frac{\eta }{x (1-x)}\right), 
\label{f0}
\\
f_1(\eta,x) &=&
-\frac{16 P_6 \big(4 x (1-x)-\eta\big)^{3/2}}{45 \eta ^{3/2} (x-1) x^3} K_1\left(\frac{\eta }{x (1-x)}\right)
+\frac{4 P_5}{45 (x-1)^3 x^5} \ln\left(\frac{\eta }{x (1-x)}\right)
\nonumber \\ &&
+\frac{32 (x-1) (4 x+1)}{3 x} K_2\left(\frac{\eta }{x (1-x)}\right)
-\frac{64 P_7}{45 x^2} \left[2 \zeta_2+\ln^2\left(\frac{\eta }{x (1-x)}\right)\right]
\nonumber \\ &&
-\frac{P_4}{45 (x-1)^3 x^5},
\label{f1}
\\
f_2(\eta,x) &=&
\frac{64 P_{10} \big(4 x (1-x)-\eta\big)^{3/2}}{9 \eta^{3/2} (x-1) x^2} K_1\left(\frac{\eta }{x (1-x)}\right)
-\frac{128}{3} (x-1) \Biggl[
K_2\left(\frac{\eta }{x (1-x)}\right)
\nonumber \\ &&
-\frac{10}{3} \ln^2\left(\frac{\eta}{x (1-x)}\right)
-\frac{20}{3} \zeta_2
\Biggr]
-\frac{4}{9 (x-1)^3 x^4} \left[4 P_9 \ln\left(\frac{\eta }{x (1-x)}\right)-P_8\right],
\label{f2}
\\
f_3(\eta,x) &=&
-\frac{16 P_{13} \sqrt{4 x (1-x)-\eta}}{9 \eta ^{3/2} (x-1) x^3} K_1\left(\frac{\eta }{x (1-x)}\right)
-\frac{32 (x-1)}{3 x} \Biggl\{
K_2\left(\frac{\eta }{x (1-x)}\right)
\nonumber \\ &&
-\frac{2}{3} (3 x+5) \left[2 \zeta_2+\ln^2\left(\frac{\eta }{x (1-x)}\right)\right]
\Biggr\}
+\frac{4 P_{12}}{9 (x-1)^3 x^5} \ln\left(\frac{\eta}{x (1-x)}\right)
\nonumber \\ &&
-\frac{P_{11}}{27 (x-1)^3 x^5},
\label{f3}
\end{eqnarray}
where the functions $K_1$ and $K_2$, which appear repeatedly in the expressions above,
are given in terms of the iterated integrals defined in Eq. (\ref{Gfunctions}),
\begin{eqnarray}
K_1(u) &=& G\left(\left\{\frac{1}{\tau },\sqrt{4-\tau } \sqrt{\tau }\right\},u\right)
+\frac{1}{2} \big(1-2 \ln (u)\big) G\left(\left\{\sqrt{4-\tau } \sqrt{\tau }\right\},u\right),
\\
K_2(u) &=& 
-G\left(\left\{\sqrt{4-\tau } \sqrt{\tau }\right\},u\right) G\left(\left\{\frac{1}{\tau },\sqrt{4-\tau } \sqrt{\tau }\right\},u\right)
+\frac{2}{3} \ln^3(u)
\nonumber \\ &&
+G\left(\left\{\frac{1}{\tau },\sqrt{4-\tau } \sqrt{\tau },\sqrt{4-\tau } \sqrt{\tau }\right\},u\right)
+4 \zeta_2 \ln(u)
+8 \zeta_3
\nonumber \\ &&
-\frac{1}{4} \big(1-2 \ln (u)\big) G^2\left(\left\{\sqrt{4-\tau } \sqrt{\tau }\right\},u\right).
\end{eqnarray}
The expressions of the $G$-functions are presented in the Appendix in terms of harmonic polylogarithms containing 
square-root valued arguments, and the $P_i$'s, with $i=1,\ldots,13$, are polynomials in $\eta$ and $x$ given by
\begin{eqnarray}
P_1 &=& 1536 (x-1)^4 (3 x+2) x^4+576 (x-1)^3 (12 x-7) \eta  x^3 \nonumber \\ &&
+8 (x-1)^2 (264 x-329) \eta ^2 x^2+16 (x-1) (12 x-37) \eta ^3 x-45 \eta ^4, \\
P_2 &=& 128 (x-1)^4 (3 x-8) x^4-32 (x-1)^3 (33 x-8) \eta  x^3 \nonumber \\ &&
-4 (x-1)^2 (108 x-133) \eta ^2 x^2-24 (x-1) (2 x-7) \eta ^3 x+15 \eta ^4, \\
P_3 &=& 4 (x-1)^2 (6 x-1) x^2-6 (x-1) (4 x+1) \eta  x+15 \eta ^2, \\
P_4 &=& 768 (x-1)^4 (40 x+7) x^4+576 (x-1)^3 (20 x-1) \eta  x^3 \nonumber \\ &&
-8 (x-1)^2 (260 x+197) \eta ^2 x^2-16 (x-1) (100 x+31) \eta ^3 x-45 (4 x+1) \eta ^4, \\
P_5 &=& 64 (x-1)^4 (40 x+13) x^4+16 (x-1)^3 (200 x+17) \eta  x^3 \nonumber \\ &&
-4 (x-1)^2 (100 x+79) \eta ^2 x^2-48 (x-1) (10 x+3) \eta ^3 x-15 (4 x+1) \eta ^4, \\
P_6 &=& 8 (x-1)^2 (10 x+1) x^2-6 (x-1) (20 x+3) \eta  x+15 (4 x+1) \eta ^2, \\
P_7 &=& 10 (x-1) x (10 x+1)-3 \eta , \\
P_8 &=& 1536 (x-1)^4 x^4+576 (x-1)^3 \eta  x^3-104 (x-1)^2 \eta ^2 x^2 \nonumber \\ &&
-80 (x-1) \eta ^3 x-9 \eta ^4, \\
P_9 &=& 128 (x-1)^4 x^4+160 (x-1)^3 \eta  x^3-20 (x-1)^2 \eta ^2 x^2-24 (x-1) \eta ^3 x-3 \eta^4, \\
P_{10} &=& 4 (x-1)^2 x^2-6 (x-1) \eta  x+3 \eta ^2, \\
P_{11} &=& 512 (x-1)^4 (7 x-9) x^4-1728 (x-1)^3 (2 x+1) \eta  x^3 \nonumber \\ &&
-24 (x-1)^2 (24 x-13) \eta ^2 x^2+240 (x-1) \eta ^3 x+27 \eta^4, \\
P_{12} &=& 32 (x-1)^4 (11 x-4) x^4-32 (x-1)^3 (6 x+5) \eta  x^3 \nonumber \\ &&
-4 (x-1)^2 (12 x-5) \eta ^2 x^2+24 (x-1) \eta ^3 x+3 \eta ^4, \\
P_{13} &=& 16 (x-1)^3 (3 x+1) x^3-4 (x-1)^2 (6 x+5) \eta  x^2+6 (x-1) \eta ^2 x+3 \eta ^3.
\end{eqnarray}

The $g_i(\eta,x)$ functions, defined in the ranges $0 <x <\eta_-$ and $\eta_+ < x <1$, are given by
\begin{eqnarray}
g_0(\eta,x) &=&
\frac{x-1}{x} \Biggl[
\frac{64 P_{15}}{45 \eta ^{3/2} x} \big(\eta -4 x (1-x)\big)^{3/2} K_3\left(\frac{x (1-x)}{\eta}\right)
+\frac{64}{3} K_4\left(\frac{x (1-x)}{\eta}\right)
\nonumber \\ &&
+\frac{32}{45 x} \left(6 \eta +30 x^2-5 x\right) \ln ^2\left(\frac{x (1-x)}{\eta}\right)
+\frac{64 \zeta_2 \left(6 \eta +30 x^2-35 x\right)}{45 x}
\nonumber \\ &&
-\frac{128 P_{14}}{45 \eta^2 x} \ln \left(\frac{x (1-x)}{\eta}\right)
+\frac{256 (x-1)}{45 \eta } \left(3 \eta +24 x^2-34 x\right)
\Biggr],
\label{g0}
\\
g_1(\eta,x) &=&
-\frac{128 P_{17}}{45 \eta ^{3/2} x^2} \big(\eta -4 x (1-x)\big)^{3/2} K_3\left(\frac{x (1-x)}{\eta}\right)
-\frac{64 P_7}{45 x^2} \ln^2\left(\frac{x (1-x)}{\eta}\right)
\nonumber \\ &&
-\frac{128 (x-1) (4 x+1)}{3 x} K_4\left(\frac{x (1-x)}{\eta}\right)
+\frac{256 P_{16}}{45 \eta^2 x^2} \ln\left(\frac{x (1-x)}{\eta}\right)
\nonumber \\ &&
+\frac{256 (x-1)}{45 \eta x} \left(3 \eta +80 x^3-36 x^2-44 x\right)
+\frac{128 \zeta_2 \left(3 \eta +20 x^3-20 x\right)}{45 x^2},
\label{g1}
\\
g_2(\eta,x) &=&
\frac{256}{3} (x-1) \Biggl[
\frac{2 \big(\eta -4 x (1-x)\big)^{3/2}}{3 \eta ^{3/2}} K_3\left(\frac{x (1-x)}{\eta}\right)
+2 K_4\left(\frac{x (1-x)}{\eta}\right)
\nonumber \\ &&
-\frac{4}{3 \eta ^2} \left(-\eta +4 x^2-4 x\right)^2 \ln \left(\frac{x (1-x)}{\eta}\right)
-\frac{16 (x-1) x}{3 \eta }
-\frac{2 \zeta_2}{3}
\nonumber \\ &&
+\frac{5}{3} \ln^2\left(\frac{x (1-x)}{\eta}\right)
\Biggr],
\label{g2}
\\
g_3(\eta,x) &=&
\frac{64 (x-1)}{x} \Biggl[
\frac{2 P_{19}}{9 \eta ^{3/2}} \sqrt{\eta-4 x (1-x)} K_3\left(\frac{x (1-x)}{\eta}\right)
+\frac{2}{3} K_4\left(\frac{x (1-x)}{\eta}\right)
\nonumber \\ &&
-\frac{2 P_{18}}{9 \eta^2} \ln\left(\frac{x (1-x)}{\eta}\right)
+\frac{8 x}{27 \eta } \left(-7 \eta +36 x^2-42 x+6\right)
+\frac{2}{9} (3 x-1) \zeta_2
\nonumber \\ &&
+\frac{1}{9} (3 x+5) \ln^2\left(\frac{x (1-x)}{\eta}\right)
\Biggr].
\label{g3}
\end{eqnarray}
Here the functions $K_3$ and $K_4$ are
\begin{eqnarray}
K_3(u) &=& G\left(\left\{\frac{1}{\tau },\frac{\sqrt{1-4 \tau }}{\tau }\right\},u\right)
-\big(\ln (u)+2\big) G\left(\left\{\frac{\sqrt{1-4 \tau }}{\tau }\right\},u\right)+\zeta_2,
\\
K_4(u) &=& 
-G\left(\left\{\frac{\sqrt{1-4 \tau }}{\tau },\frac{\sqrt{1-4 \tau }}{\tau },\frac{1}{\tau }\right\},u\right)
+\zeta_2 G\left(\left\{\frac{\sqrt{1-4 \tau }}{\tau }\right\},u\right)
\nonumber \\ &&
-2 G\left(\left\{\frac{\sqrt{1-4 \tau }}{\tau },\frac{\sqrt{1-4 \tau }}{\tau }\right\},u\right)
+\zeta_2 \ln (u)
+\frac{1}{6} \ln ^3(u),
\end{eqnarray}
and
\begin{eqnarray}
P_{14} &=& 16 (x-1)^2 (6 x-1) x^3-8 (x-1) (9 x-4) \eta  x^2-(36 x-41) \eta ^2 x-6 \eta ^3, \\
P_{15} &=& x (6 x-1)-6 \eta , \\
P_{16} &=& 32 (x-1)^3 (10 x+1) x^3-4 (x-1)^2 (40 x+1) \eta  x^2 \nonumber \\ &&
+(x-1) (20 x+23) \eta ^2 x+3 \eta ^3, \\
P_{17} &=& 2 (x-1) x (10 x+1)+3 \eta , \\
P_{18} &=& 32 (x-1)^2 (3 x+1) x^2-16 (x-1) (3 x+1) \eta  x+(2-7 x) \eta ^2, \\
P_{19} &=& 4 (x-1) x (3 x+1)+(1-6 x) \eta.
\end{eqnarray}

Finally, the $h_i(\eta,x)$ functions, defined in the full range $0 < x < 1$, are just given by the $g_i(\eta,x)$ functions with $\eta \rightarrow 1/\eta$, i.e.,
\begin{equation}
h_i(\eta,x) = g_i\left(\frac{1}{\eta},x\right), \quad i=0,1,2,3.
\end{equation}

We see that iterated integrals of up to weight three appear in our result. The alphabet of these integrals is given in terms of just three letters:
\begin{equation}
\frac{1}{\tau}, \quad \sqrt{4-\tau} \sqrt{\tau}, \quad \frac{\sqrt{1-4 \tau}}{\tau}.
\end{equation}
In principle, we could try to calculate all of the integrals in $y$ appearing in Eq. (\ref{aQq}) and express them in terms of iterated integrals of 
higher weight. However, this is not really necessary or even convenient, since the expressions (\ref{f0}--\ref{f3}, \ref{g0}--\ref{g1}) are very compact,
and integrating them into higher weight iterated integrals leads to a result of considerably larger size. 
Furthermore, all of the iterated integrals appearing above can be written in terms of simple polylogarithms (albeit of complicated arguments), 
see the Appendix, for which various fastly converging numerical representations exist. Therefore, the integrals in $y$ 
appearing in Eq. (\ref{aQq}) can be performed 
numerically without problems. The convolution with parton distribution functions (which will be presented in a 
future phenomenological publication), in order to compute the corresponding contribution to $F_2(x,Q^2)$ or for the 
transition rate in the VNFS, is straightforward.
\section{Numerical Results}
\label{sec:5}

\begin{figure}[H]\centering
\includegraphics[width=0.7\textwidth]{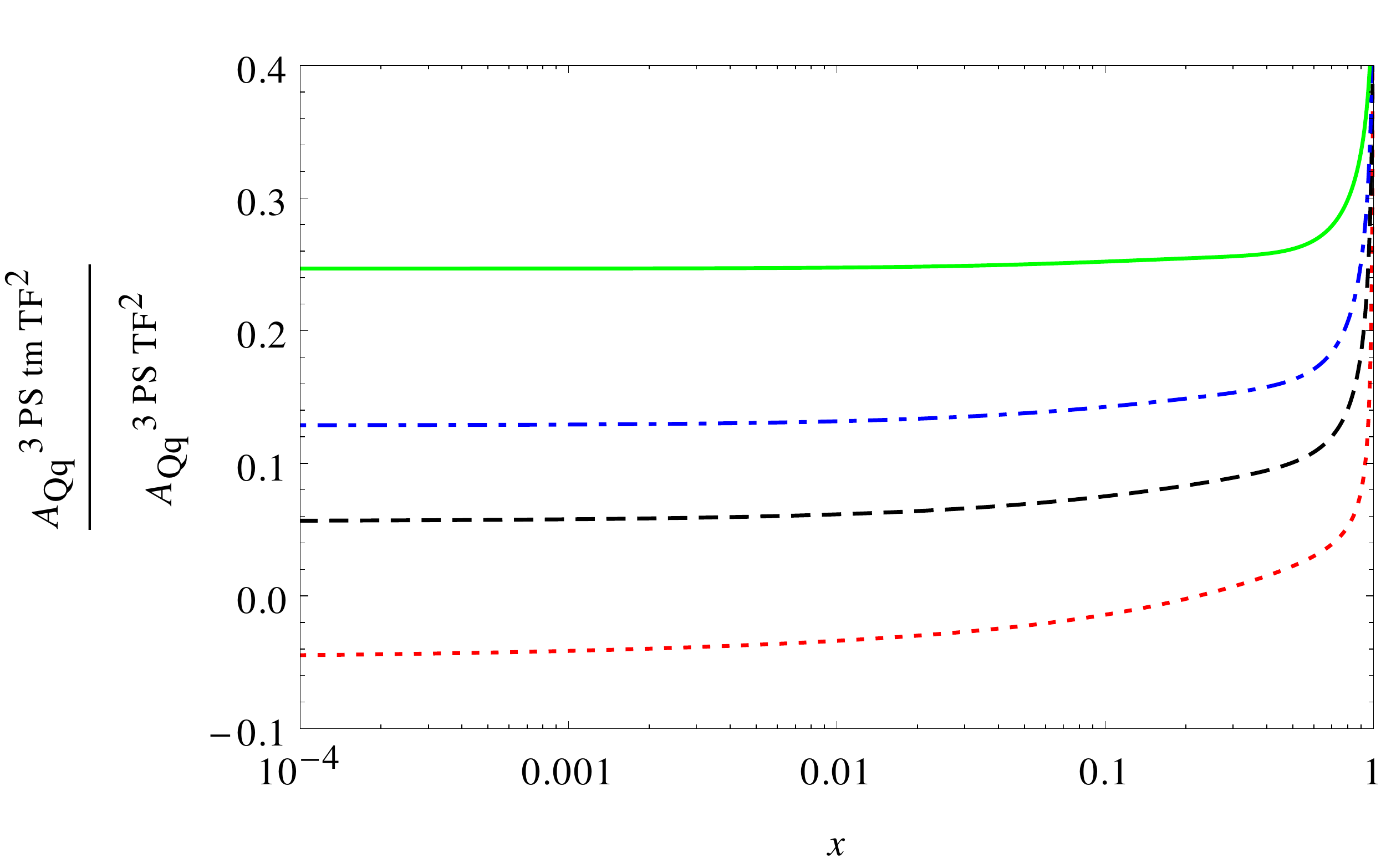}
\caption[]{\label{FIGnum} \sf 
The ratio of the 2-mass (tm) contributions to the massive OME $A_{Qq}^{{\sf PS},(3)}$ to all contributions to 
$A_{Qq}^{{\sf PS},(3)}$
of $O(T_F^2)$ as a function of $x$ and $\mu^2$. 
Dotted line (red): $\mu^2 = 30~\GeV^2$.
Dashed line (black): $\mu^2 = 50~\GeV^2$.
Dash-dotted line (blue): $\mu^2 = 100~\GeV^2$.
Full line (green): $\mu^2 = 1000~\GeV^2$. Here the on-shell heavy quark masses $m_c = 1.59~\GeV$ and $m_b = 
4.78~\GeV$ 
\cite{Alekhin:2012vu,Agashe:2014kda} have been used.
}
\end{figure}

\vspace*{1mm}
\noindent
We compare the pure singlet 2-mass contributions to the  complete $O(T_F^2 C_{A,F})$ term  
as a function of $x$ and $\mu^2$ in Figure~\ref{FIGnum}.
\noindent
Typical virtualities are $\mu^2 \in [30,1000]~\GeV^2$. The ratio of the 2-mass contributions to the complete term 
of $O(T_F^2 C_{A,F})$ grows in this region from slightly negative contributions to $\sim 0.36$ for very large
virtualities in most of the $x$-range. The behaviour of the ratio is widely flat 
in $x$, rising at very large $x$. 
\section{Conclusions}
\label{sec:6}

\vspace*{1mm}
\noindent
We have calculated the two-mass 3-loop contributions to the massive OME $A_{Qq}^{{\sf PS},(3)}$ in analytic form in 
$x$-space for a general mass-ratio $\eta$. It contributes
to the 3-loop VFNS and to the massive 3-loop corrections of the deep-inelastic structure function $F_2(x,Q^2)$
in the region $m^2 \ll Q^2$.  As a function of $x$ its relative contribution to the $O(T_F^2 C_{A,F})$ terms of the 
whole matrix element $A_{Qq}^{{\sf PS},(3)}$ behaves widely flat and grows with the scale $\mu^2$ up to about $\sim 
0.36$.

We used Mellin-Barnes techniques to obtain the $x$-space result by factoring out the $N$-dependence in terms of the 
kernel $x^N$, and used integration by parts to absorb the $N$-dependent polynomial pre-factors. The result can be 
written as single limited integrals within the range $x \in [0,1]$ over iterated integrals containing also square-root 
valued letters. These integrals can be turned into polylogarithms of involved root-valued arguments, depending on 
$\eta$. The even Mellin moments of the OME exhibit a growing number of polynomial terms in $\eta$ with growing values 
of $N$. Due to this structural property and the arbitrariness of $\eta$, which enters the ground field, the method of 
arbitrarily large moments \cite{Blumlein:2017dxp} cannot be used to find the result in the present case. 
\appendix
\section{\boldmath Evaluating the $G$-functions}
\label{sec:7}

\subsection{\boldmath $G$-functions with support $0<x<1$}

\vspace*{1mm}
\noindent
Here, we have the argument 
\begin{eqnarray}
\xi_1 = x(1-x) \eta \in ( 0 , \eta/4 )
\end{eqnarray}
and therefore $\sqrt{ 4 \xi_1 - 1 } = i \sqrt{ 1 - 4 \xi_1 }$.
We also define 
\begin{eqnarray}
\omega_1 = \sqrt{1-4\xi_1}.
\end{eqnarray}

We obtain:
\begin{eqnarray}
    G \left( \left\{ \frac{\sqrt{1 - 4 \tau}}{\tau } \right\}, \xi_1 \right) 
    &=&  2  \omega_1 + 2 \ln \big( 1 - \omega_1 \big) - \ln (4 \xi_1 ) - 2 \\
    G \left( \left\{ \tau  \left| 1-4 \tau \right| \right\} , \xi_1 \right)   
    &=& \xi_1^2 \left( \frac{1}{2} - \frac{4 \xi_1}{3} \right) \\
    G \left( \left\{ \tau ^2 \left| 1-4 \tau \right| \right\} , \xi_1 \right) 
    &=& \xi_1^3 \left( \frac{1}{3}- \xi_1 \right) \\
    G \left( \left\{ \frac{1}{\tau} , \frac{ \sqrt{1 - 4 \tau} }{\tau} \right\} , \xi_1\right) 
    &=& 4 \omega_1 - \ln^2 \big( 1 - \omega_1 \big) -\frac{1}{2}  \ln^2 \big( 4 \xi_1 \big) \NN \\ 
    && + 4 \ln \big( 1 - \omega_1 \big) - 2 \ln (2) \ln \big( 1 - \omega_1 \big)  \NN \\
    && - 4 \ln \big( 4 \xi_1 \big) + 2 \ln \big( 1 - \omega_1 \big) \ln \big( 4 \xi_1\big) \NN \\
    && + 2 \text{Li}_2\left( \frac{1 -\omega_1}{2} \right) - 4 + \ln ^2(2) + 4 \ln (2) \\
    G \left( \left\{ \frac{ \sqrt{1 - 4 \tau}}{\tau } ,\frac{1}{\tau} \right\} , \xi_1 \right) 
    &=& \omega_1 \left( -4 -4 \ln (2) +2  \ln \big(4 \xi_1\big) \right) \NN \\
    && + \ln ^2 \left( 1 - \omega_1 \right) - \frac{1}{2} \ln ^2 \left( 4 \xi_1 \right) - 4  \ln \big( 1 - \omega_1 \big) \NN \\
    && -2  \ln (2) \ln \big( 1 - \omega_1 \big) + 2 \ln \big( 4 \xi_1 \big) \NN \\
    && + 2 \ln (2) \ln \big( 4 \xi_1 \big) -2  \text{Li}_2\left( \frac{1-\omega_1}{2} \right) + 4 - \ln ^2(2) \\
    G \left( \left\{ \frac{\sqrt{1 - 4 \tau}}{\tau } , \frac{\sqrt{1 - 4 \tau }}{\
\tau } \right\} , \xi_1 \right) 
    &=&  \omega_1 \left(  4 \ln \big( 1 - \omega_1 \big) - 4 - 2 \ln \big(4 \xi_1\big) \right)
- 8 \xi_1 \NN \\
    && + 2 \ln ^2 \left( 1 - \omega_1 \right) + \frac{1}{2} \ln ^2 \left( 4 \xi_1 \right) - 4 \ln\left( 1 - \omega_1 \right) \NN \\ 
    && + 2 \ln \left( 4 \xi_1 \right) - 2 \ln \left( 1 - \omega_1 \right) \ln \left( 4 \xi_1 \right) + 4 \\
    G \left( \left\{ \frac{1}{\tau} , \frac{1}{\tau} , \frac{\sqrt{1 - 4 \tau}}{\tau } \right\} , \xi_1 \right) 
    &=& 2 \text{Li}_3 \left( \frac{\omega_1 + 1}{\omega_1 - 1} \right) -4 \ln \left( 1 - \omega_1\right) \zeta_2 + 2 \ln \left(4 \xi_1 \right) \zeta_2 \NN \\
&& + 8 \omega_1 - 2 \ln^3 \left( 1 - \omega_1 \right) - \frac{1}{6} \ln^3 \left( 4 \xi_1 \right) + 2 \ln^2 (2) \NN \\
&& \times \ln \left( 1 - \omega_1 \right) - 2 \ln^2 \left( 1 - \omega_1 \right) + 2 \ln (2) \NN \\
&& \times \ln^2 \left( 1-\omega_1 \right) + 2 \ln^2 \left( 1 - \omega_1 \right) \ln \left( 4 \xi_1 \right) - 2 \ln^2 
\left( 4 \xi_1 \right) \NN \\
&& + \ln (2) \ln^2 \left( 4 \xi_1 \right) + 8 \ln \left( 1 - \omega_1 \right) - 4 \ln (2) \NN \\
&& \times \ln \left(1-\omega_1\right) -8 \ln \left(4 \xi_1 \right) + 4 \ln (2) \ln \left(4 \xi_1\right)  \NN \\ 
&& + 4 \ln \left( 1-\omega_1\right) \ln \left(4 \xi_1\right) - 4 \ln (2) \ln \left(1-\omega_1\right) \ln \left(4 \xi_1\right) \NN \\
&& + 4 \text{Li}_2\left(\frac{1 - \omega_1}{2} \right) + 4 \text{Li}_3 \left(\frac{1 - \omega_1}{2} \right) - 8 - \frac{2}{3} \ln^3 (2) \NN \\
&& -2 \ln^2 (2) + 8 \ln (2) \\
    G \left( \left\{ \frac{1}{\tau} , \frac{\sqrt{1 - 4 \tau}}{\tau} , \frac{1}{\tau} \right\} , \xi_1 \right) 
    &=& \left( - 16 - 4 \ln \left( 4 \xi_1 \right) + 2 \ln^2 \left( 4 \xi_1 \right) + 8 \zeta_2 \right) \ln \left(1-\omega_1 \right)  \NN \\
&& + \ln^2 \left( 1-\omega_1 \right) \left( 4 - 5 \ln \left( 4 \xi_1 \right) \right) + \Bigl( 12 \NN \\
&& + 2 \text{Li}_2\left(\frac{1 - \omega_1}{2} \right) + 4 \omega_1 - 4 \zeta_2 \Bigr) \ln \left( 4 \xi_1 \right) \NN \\
&& - 4 \Bigl( - 4 + 2 \text{Li}_2 \left( \frac{1 - \omega_1}{2} \right) + 2 \text{Li}_3 \left(\frac{1 - \omega_1}{2} \right) \NN \\
&& + \text{Li}_3 \left( \frac{\omega_1+1}{\omega_1-1}\right) + 4 \omega_1 \Bigr) + \Bigl( -8 - 4 \text{Li}_2 \left( \frac{1 - \omega_1}{2} \right) \NN \\
&& - 8 \omega_1 - 2 \ln^2 \left( 1 - \omega_1 \right) + 4 \ln \left(4 \xi_1\right)+ 2 \ln \left(1-\omega_1\right) \NN \\
&& \times \ln \left(4 \xi_1\right) - \ln^2 \left(4 \xi_1\right) \Bigr) \ln (2) + 4 \ln^3 \left( 1 - \omega_1 \right) \NN \\
&& - \frac{1}{6} \ln^3 \left( 4 \xi_1 \right) + \ln^2 (2) \left( \ln \left(4 \xi_1\right) - 4 \right) - \frac{2}{3}  \ln^3 (2) \\
    G \left( \left\{ \frac{1}{\tau} , \frac{\sqrt{1 - 4 \tau}}{\tau } , \frac{1 - \sqrt{4 \tau}}{\tau } \right\} , \xi_1 \right)
    &=& - \left( \ln \left( 1-\omega_1 \right) \left( 2 \ln \left( 4 \xi_1 \right) - 4 \right) -\ln^2 \left( 4 \xi_1 \right) \right) \ln (2) \NN \\
&& - \left( 2 - 2 \ln \left( 1 - \omega_1 \right) + \ln \left( 4 \xi_1 \right) \right) \ln^2 (2) - \ln^2 \left( 1 - \omega_1 \right) \NN \\
&& \times \left( - 5 \ln \left( 4 \xi_1 \right) - 6 \right) - \Bigl( 8 - 4 \text{Li}_2 \left( \frac{1-\omega_1}{2} \right) - 8 \omega_1 \NN \\
&& + 8 \ln \left( 4 \xi_1 \right) + 2 \ln^2 \left( 4 \xi_1 \right) + 4 \zeta_2\Bigr) \ln \left( 1 - \omega_1 \right) \NN \\
&& - \left(-4 + 2\text{Li}_2 \left( \frac{1-\omega_1}{2} \right) + 4 \omega_1 - 2 \zeta_2 \right) \ln \left(4 \xi_1\right) \NN \\
&& - 2 \left( - 4 - \text{Li}_3 \left( \frac{1 + \omega_1}{\omega_1-1} \right) 
+ 2 \text{Li}_2\left( \frac{1 - \omega_1}{2} \right) + 4 \omega_1 + 4 \xi_1 \right) \NN \\
&& - \frac{10}{3} \ln^3 \left( 1- \omega_1 \right) + \frac{1}{6} \ln^3 \left(4 \xi_1\right) + 2 \ln^2 \left( 4 \xi_1 \right) \\
    G \left( \left\{ \frac{\sqrt{1 - 4 \tau}}{\tau },\frac{1}{\tau },\frac{1}{\tau } \right\} , \xi_1 \right)
    &=& 2 \Bigl( - 4 + 2 \text{Li}_2 \left( \frac{1-\omega_1}{2} \right) \NN \\
&& + 2 \text{Li}_3 \left( \frac{1 - \omega_1}{2} \Bigr) 
+ \text{Li}_3 \left( \frac{1+\omega_1}{\omega_1-1} \right) + 4 \omega_1 \right) \NN \\
&& + \Bigl( 4 \text{Li}_2 \left( \frac{1-\omega_1}{2} \right) + 8 \omega_1 
+ \ln \left( 1 - \omega_1 \right) \left( 4 - 2  \ln \left( 4 \xi_1 \right) \right) \NN \\
&& - 4 \left( 1 + \omega_1 \right) \ln \left(4 \xi_1\right) + 2 \ln^2 \left(4 \xi_1\right) \Bigr) \ln (2) \NN \\ 
&& + \left( 2 \left( 2 \omega_1 + 1 \right) + 2 \ln \left( 1 - \omega_1 \right) - 3 \ln \left(4 \xi_1\right)\right) \ln^2 (2) \NN \\
&& + \left( 8 - \ln^2 \left( 4 \xi_1 \right) - 4 \zeta_2 \right) \ln \left( 1 - \omega_1 \right)  + \ln^2 \left( 1 - \omega_1 \right) \NN \\
&& \times \Bigl( 3 \ln \left( 4 \xi_1 \right) - 2 \Bigr) + \Bigl( -4 - 2\text{Li}_2 \left( \frac{1-\omega_1}{2} 
\right) - 4 \omega_1 \NN \\
&& + 2 \zeta_2\Bigr) \ln \left(4 \xi_1\right) + \frac{4}{3} \ln^3 (2) - 2 \ln^3 \left( 1 - \omega_1 \right) \NN \\
&& + \left( 1 + \omega_1 \right) \ln^2 \left(4 \xi_1\right) - \frac{1}{6}  \ln^3 \left(4 \xi_1\right) \\
    G \left( \left\{ \frac{\sqrt{1 - 4 \tau}}{\tau },\frac{1}{\tau},\frac{\sqrt{1 - 4 \tau}}{\tau } \right\} , 
    \xi_1 \right) 
    &=& 4 \Bigl( - \text{Li}_3 \left( \frac{\omega_1 + 1}{\omega_1 - 1} \right) + \text{Li}_2 \left( \frac{1-\omega_1}{2} \right) + \text{Li}_2 \left( \frac{1-\omega_1}{2} \right) \NN \\
&& \times \omega_1 - 4 \xi_1 \Bigr) - \Bigl( \left( 4 \left( \omega_1 - 1 \right) - 6 \ln \left(4 \xi_1\right)\right) \ln \left( 1-\omega_1 \right) \NN \\
&& - 8 \left( \omega_1 - 1 \right) + 4 \ln^2 \left( 1 - \omega_1 \right) + 4 \ln \left( 4 \xi_1 \right) + 2 \ln^2 \left(4 \xi_1\right)\Bigr) \NN \\
&& \times \ln (2)  - \left( - 2 \left( 1 + \omega_1 \right) + 2 \ln \left( 1 - \omega_1 \right) - \ln \left( 4 \xi_1 \right)\right) \ln^2 (2) \NN \\
&& - \Bigl( - \ln^2 \left( 4 \xi_1 \right) + 4 \text{Li}_2 \left( \frac{1-\omega_1}{2} \right) - 8 \zeta_2 \NN \\
&& - 4 \ln \left( 4 \xi_1 \right) \omega_1 \Bigr) \ln \left( 1 - \omega_1 \right)  - \Bigl( 2 \left( 1 + \omega_1 \right) \NN \\
&& + 5 \ln \left( 4 \xi_1 \right) \Bigr) \ln^2 \left( 1-\omega_1 \right) - \Bigl( -4 - 2 \text{Li}_2 \left( \frac{1 - \omega_1}{2} \right) \NN \\
&& + 4 \omega_1 + 4 \zeta_2 \Bigr) \ln \left(4 \xi_1\right) + \frac{14}{3} \ln^3 \left( 1-\omega_1 \right) \NN \\
&& - \left( \omega_1 - 1 \right) \ln^2 \left(4 \xi_1\right) + \frac{1}{6} \ln^3 \left(4 \xi_1\right)  \\
    G \left( \left\{ \frac{\sqrt{1 - 4 \tau}}{\tau }, \frac{\sqrt{1 - 4 \tau}}{\tau } , \frac{1}{\tau} \right\} , \xi_1 \right) 
    &=&
\Bigl(
        8
        -4 \ln ^2(2)
        -2 \zeta_2
        -4 \ln (2) \omega_1
\Bigr) \ln \big(4 \xi_1\big)
\NN \\ &&
+ \Biggl[-4 \big(
                1
                +2 \xi_1
                +\omega_1
        \big)
        +6 \ln ^2(2)
        -4 \ln \big(
                4 \xi_1\big)
\NN \\ &&
        +\ln ^2\big(
                4 \xi_1\big)
        +4 \ln ^2\big(
                \omega_1+1\big)
        +4 \zeta_2
        +\ln \big(
                1 + \omega_1
        \big)
\NN \\ && \times
\big(-4 \omega_1-8 \ln (2)\big)
        +4 \ln (2) \big(
                1 + \omega_1\big)
\Biggr] \ln \big(
        1-\omega_1\big)
\NN \\ && 
+\Bigl(
        4 \big(
                -3
                -2 \xi_1
                +\omega_1
        \big)
        +10 \ln ^2(2)
        +4 \ln (2) \big(
                2 \omega_1+1\big)
\Bigr) 
\NN \\ &&  \times
\ln \big(
        1 + \omega_1\big)
+\Bigl(
        -4 \ln (2)
        +4 \ln \big(
                \omega_1+1\big)
        -4 \omega_1
\Bigr) 
\NN \\ && \times
\text{Li}_2\left(
        \frac{1-\omega_1}{2} \right)
+4 \zeta_2
+4 \zeta_3
+8 \big(
        3 \xi_1
        +\omega_1
        -1
\big)
\NN \\ &&
+\Bigl(
        2
        -2 \ln \big(
                4 \xi_1
        \big)
-2 \ln (2)\Bigr) \ln ^2\big(
        1-\omega_1\big)
+\ln ^2\big(
        4 \xi_1
\big)
\NN \\ && \times
\Bigl(\omega_1+2 \ln (2)+1\Bigr)
-\frac{1}{2} \ln ^3\big(
        4 \xi_1\big)
+16 \ln (2) \xi_1
\NN \\ &&
-4 \text{Li}_2\left(
        \frac{1 - \omega_1}{2} \right)
-2 \text{Li}_3\left(
        \frac{\omega_1+1}{\omega_1-1}\right)
-4 \text{Li}_3\left(
        \frac{1-\omega_1}{2} \right)
\NN \\ &&
-4 \text{Li}_3\left(
        \frac{1 + \omega_1}{2} \right)
+\Bigl(
        4 \ln \big(
                1
                +\omega_1
        \big)
-4 \ln (2)-4\Bigr) 
\NN \\ &&
\text{Li}_2\left(
        \frac{1 + \omega_1}{2} \right)
+2 \ln ^3\big(
        1-\omega_1\big)
+\frac{2}{3} \ln ^3\big(
        1 + \omega_1\big)
\NN \\ &&
-\Bigl(
        2 \big(
                1
                +\omega_1
        \big)
+6 \ln (2)\Bigr) \ln ^2\big(
        \omega_1+1\big)
-2 \ln ^2(2) 
\NN \\ && \times
\big(
        \omega_1+2\big)
-\frac{8}{3} \ln ^3(2).
\end{eqnarray}

\subsection{\boldmath $G$-functions with support $x \in [0 ,\frac{1}{2}-\frac{1}{2} \sqrt{1-\eta}] \ \cup \ 
[\frac{1}{2}+\frac{1}{2} \sqrt{1-\eta},1]$}

The integrals in this class coincide with integrals with full support ($0<x<1$), but the replacements 
\begin{align}
    \xi_1 & \to \xi_3 = \frac{x(1-x)}{\eta} ,\\
    \omega_1 &\to \omega_3 = \sqrt{ 1 - 4 \xi_3}
\end{align}
have to be performed.

\subsection{\boldmath $G$-functions with support $\frac{1}{2}-\frac{1}{2} \sqrt{1-\eta} < x < \frac{1}{2}+\frac{1}{2} 
\sqrt{1-\eta}$}

Here, we have the argument 
\begin{equation}
\xi_2 = \frac{\eta}{x(1-x)} \in ( 4 \eta , 4  )
\end{equation} 
and therefore $\sqrt{ 4 - \xi_2 }$ is real.
We introduce the abbreviations 
\begin{eqnarray}
\omega_2 &=& \sqrt{\xi_2} \sqrt{4-\xi_2}, \\
\phi &=& \arcsin \left( \frac{\sqrt{\xi_2}}{2} \right).
\end{eqnarray} 

\begin{eqnarray}
G \left( \left\{ \sqrt{ 4 - \tau} \sqrt{\tau } \right\} , \xi_2 \right)
&=& - \omega_2 \left( 1 - \frac{\xi_2}{2} \right) + 4  \phi  \\
G \left( \left\{ \frac{1}{4-\tau },\sqrt{4-\tau } \sqrt{\tau} \right\} , \xi_2 \right)
&=& - \frac{1}{2} \omega_2 \left( 1 + \frac{\xi_2}{2} \right) + \Bigl( 2 - 4 \ln ( 4 - \xi_2) \Bigr)  \phi \NN \\
&& + 4 \text{Cl}_2 \left( 2 \phi \right) - 2 \text{Cl}_2 \left( 4 \phi \right) \\
G \left( \left\{ \frac{1}{\tau },\sqrt{4-\tau } \sqrt{\tau} \right\} , \xi_2 \right) 
&=& \omega_2 \left( \frac{\xi_2}{4} - \frac{3}{2} \right) - \Bigl( 2 - 4 \ln ( \xi_2 ) 
\Bigr) \phi + 4 \text{Cl}_2 \left( 2 \phi \right) 
\nonumber\\
\\
G \left( \left\{ \frac{1}{\tau },\sqrt{4-\tau } \sqrt{\tau},\sqrt{4-\tau} \sqrt{\tau }\right\}, \xi_2 \right)
&=& 5 \xi_2 - \frac{3}{2} \xi_2^2 + \frac{1}{3} \xi_2^3 - \frac{1}{32} \xi_2^4 - 8 \zeta_3 + \phi \Bigl[ \left( \xi_2 - 6 \right) \omega_2 \NN \\
&& + 16 \text{Cl}_2 \left( 2 \phi \right) \Bigr] + 8 \text{Cl}_3 \left( 2 \phi \right) + \phi^2 \left( 8 \ln ( \xi_2 ) 
- 4 \right). \nonumber\\ 
\end{eqnarray}

We used the Clausen function \cite{LEWIN1,LEWIN2}
\begin{equation}
\begin{split}
\text{Cl}_{2} (x) &= \frac{i}{2} \left( \text{Li}_2 ( e^{-ix} ) - \text{Li}_2 ( e^{ix} ) \right), \\
\text{Cl}_{3} (x) &= \frac{1}{2} \left( \text{Li}_3 ( e^{-ix} ) + \text{Li}_3 ( e^{ix} ) \right) \\
\end{split}
\end{equation}
with the sum representation
\begin{equation}
\begin{split}
\text{Cl}_2 ( \phi ) &= \sum\limits_{n=1}^\infty \frac{ \sin \left( n \phi \right) }{n^2}, \\
\text{Cl}_3 ( \phi ) &= \sum\limits_{n=1}^\infty \frac{ \cos \left( n \phi \right) }{n^3}
\end{split}
\end{equation}
for $\phi \in (0,2\pi)$.


\section{Fixed Moments}
For fixed values of $N = 2k, k \in \mathbb{N} \backslash \{0\}$, we find the following moments:

\begin{eqnarray}
\lefteqn{A_{Qq}^{{\sf PS},(3)} \left( N=2 \right)= } \NN \\ &&
-\frac{8192}{81 \ep^3}
+\frac{1}{\ep^2}
\Bigl[
-\frac{11776}{243}
-\frac{2048}{27} \left( L_2 + L_1 \right)
\Bigr]
+\frac{1}{\ep}
\Bigl[
-\frac{75136}{729}
-\frac{512}{9} \left( L_2^2 + L_1^2 \right) \NN \\ &&
-\frac{2944}{81} \left( L_2 + L_1 \right)
-\frac{512}{9} H_0(\eta ) \left( L_2 - L_1 \right)
-\frac{1024}{27} H_0^2(\eta )
-\frac{1024}{27} \zeta_2
\Bigr]
-\frac{125600}{2187} \NN \\ &&
-\frac{256}{9} \left( L_2^3 + L_1^3 \right)
-\frac{736}{27} \left( L_2^2 + L_1^2 \right)
-\frac{128}{3} H_0(\eta ) \left( L_2^2 - L_1^2 \right)
-\Bigl(
         \frac{18784}{243}
        +\frac{256}{9} H_0^2(\eta ) \NN \\ &&
        +\frac{256}{9} \zeta_2
\Bigr) \left( L_2 + L_1 \right)
-\frac{640}{27} H_0(\eta ) \left( L_2 - L_1 \right)
+\left(
        -\frac{496}{81}
        +\frac{256}{27} H_1(\eta )
\right) H_0^2(\eta ) \NN \\ &&
+\frac{128}{81} H_0^3(\eta )
-\frac{512}{27} H_0(\eta ) H_{0,1}(\eta )
+\frac{512}{27} H_{0,0,1}(\eta )
-\frac{1472}{81} \zeta_2
+\frac{1024}{81} \zeta_3 \NN \\ &&
+\left(
        \frac{320}{27}
        +\frac{40}{27} H_0^2(\eta )
\right) \left( \eta + \frac{1}{\eta} \right)
+\frac{160}{27} H_0(\eta ) \left( \eta - \frac{1}{\eta} \right)
+\Bigl(
    \left(
        32 H_{0,-1}\big(\sqrt{\eta }\big)
        +8 H_{0,1}(\eta )
    \right) \NN \\ &&
\times H_0(\eta )
-4 \left(
     H_{-1}\big(\sqrt{\eta }\big)
    + H_1\big(\sqrt{\eta }\big)
\right) H_0^2(\eta )
-64 H_{0,0,-1}\big(\sqrt{\eta }\big)
-8 H_{0,0,1}(\eta )
\Bigr) \left( \sqrt{\eta} + \frac{1}{\sqrt{\eta}} \right) \NN \\ &&
+\Biggl[
\left(
        \frac{160}{27} H_{0,-1}\big(\sqrt{\eta }\big)
        +\frac{40}{27} H_{0,1}(\eta )
\right) H_0(\eta )
-\frac{20}{27} \big(
         H_{-1}\big(\sqrt{\eta }\big)
        + H_1\big(\sqrt{\eta }\big)
\big) H_0^2(\eta ) \NN \\ &&
-\frac{320}{27} H_{0,0,-1}\big(\sqrt{\eta }\big)
-\frac{40}{27} H_{0,0,1}(\eta )
\Biggr] \left( \eta^{3/2} + \frac{1}{\eta^{3/2}} \right)
\end{eqnarray}


\begin{eqnarray}
\lefteqn{A_{Qq}^{{\sf PS},(3)}  \left( N=4 \right) = } \NN \\ && 
-\frac{30976}{2025 \ep^3}
+\frac{1}{\ep^2}
\Bigl[
-\frac{17888}{6075}
-\frac{7744}{675} \left( L_2 + L_1 \right)
\Bigr]
+\frac{1}{\ep}
\Bigl[
-\frac{6600284}{455625}
-\frac{1936}{225} \left( L_2^2 + L_1^2 \right) \NN \\ &&
-\frac{4472}{2025} \left( L_2 + L_1 \right)
-\frac{1936}{225} H_0(\eta ) \left( L_2 - L_1 \right)
-\frac{3872}{675} H_0^2(\eta )
-\frac{3872}{675} \zeta_2
\Bigr]
-\frac{24497203}{5467500} \NN \\ &&
-\frac{968}{225} \left( L_2^3 + L_1^3 \right)
-\frac{1118}{675} \left( L_2^2 + L_1^2 \right)
-\frac{484}{75} H_0(\eta ) \left( L_2^2 - L_1^2 \right)
-\Biggl[
        \frac{1650071}{151875}
        +\frac{968}{225} H_0(\eta )^2 \NN \\ &&
        +\frac{968 \zeta_2}{225}
\Biggr] \left( L_2 + L_1 \right)
-\frac{4294}{3375} H_0(\eta ) \left( L_2 - L_1 \right)
+\left(
        \frac{186109}{324000}
        +\frac{968}{675} H_1(\eta )
\right) H_0^2(\eta ) \NN \\ &&
+\frac{484}{2025} H_0^3(\eta )
-\frac{1936}{675} H_0(\eta ) H_{0,1}(\eta )
+\frac{1936}{675} H_{0,0,1}(\eta )
-\frac{2236}{2025} \zeta_2
+\frac{3872}{2025} \zeta_3 \NN \\ &&
+\left(
        \frac{1322}{675}
        +\frac{1273}{5400} H_0^2(\eta )
\right) \left( \eta + \frac{1}{\eta} \right)
+\frac{5239}{5400} H_0(\eta ) \left( \eta - \frac{1}{\eta} \right)
+\left(
        -\frac{49}{200}
        -\frac{49}{1600} H_0^2(\eta )
\right) \NN \\ &&
\times \left( \eta^2 + \frac{1}{\eta^2} \right) 
-\frac{49}{400} H_0(\eta ) \left( \eta^2 - \frac{1}{\eta^2} \right)
+\Biggl[
        \left(
                \frac{39}{8} H_{0,-1}\big(\sqrt{\eta }\big)
                +\frac{39}{32} H_{0,1}(\eta )
        \right) H_0(\eta ) \NN \\ &&
         -\frac{39}{64} \big(
                 H_{-1}\big(\sqrt{\eta }\big)
                + H_1\big(\sqrt{\eta }\big)
        \big) H_0^2(\eta )
        -\frac{39}{4} H_{0,0,-1}\big(\sqrt{\eta }\big)
        -\frac{39}{32} H_{0,0,1}(\eta )
\Biggr] \left( \sqrt{\eta} + \frac{1}{\sqrt{\eta}} \right) \NN \\ &&
+\Biggl[
        \left(
                \frac{425}{432} H_{0,-1}\big(\sqrt{\eta }\big)
                +\frac{425}{1728} H_{0,1}(\eta )
        \right) H_0(\eta )
        -\frac{425}{3456} \big(
                 H_{-1}\big(\sqrt{\eta }\big)
                + H_1\big(\sqrt{\eta }\big)
        \big) H_0^2(\eta ) \NN \\ &&
        -\frac{425}{216} H_{0,0,-1}\big(\sqrt{\eta }\big)
        -\frac{425}{1728} H_{0,0,1}(\eta )
\Biggr] \left( \eta^{3/2} + \frac{1}{\eta^{3/2}} \right)
+\Bigl(
        \Bigl(
                -\frac{49}{400} H_{0,-1}\big(\sqrt{\eta }\big) \NN \\ &&
                -\frac{49}{1600} H_{0,1}(\eta )
        \Bigr) H_0(\eta )
        + \frac{49}{3200} \big(
                 H_{-1}\big(\sqrt{\eta }\big)
                + H_1\big(\sqrt{\eta }\big)
        \big) H_0^2(\eta )
        +\frac{49}{200} H_{0,0,-1}\big(\sqrt{\eta }\big) \NN \\ &&
        +\frac{49}{1600} H_{0,0,1}(\eta )
\Bigr) \left( \eta^{5/2} + \frac{1}{\eta^{5/2}} \right)
\end{eqnarray}


\begin{eqnarray}
\lefteqn{A_{Qq}^{{\sf PS},(3)} \left( N=6 \right) =} \NN \\ && 
-\frac{123904}{19845 \ep^3}
+\frac{1}{\ep^2}
\Bigl[
-\frac{121472}{297675}
-\frac{30976}{6615} \left( L_2 + L_1 \right)
\Bigr]
+\frac{1}{\ep}
\Bigl[
-\frac{257649488}{43758225}
-\frac{7744}{2205} \left( L_2^2 + L_1^2 \right) \NN \\ &&
-\frac{30368}{99225} \left( L_2 + L_1 \right)
-\frac{7744}{2205} H_0(\eta ) \left( L_2 - L_1 \right)
-\frac{15488}{6615} H_0^2(\eta )
-\frac{15488}{6615} \zeta_2
\Bigr]
-\frac{18655921961}{17503290000} \NN \\ &&
-\frac{3872}{2205} \left( L_2^3 + L_1^3 \right)
-\frac{7592}{33075} \left( L_2^2 + L_1^2 \right)
-\frac{1936}{735} H_0(\eta ) \left( L_2^2 - L_1^2 \right)
+\Bigl(
        -\frac{64412372}{14586075} \NN \\ &&
        -\frac{3872}{2205} H_0^2(\eta )
        -\frac{3872}{2205} \zeta_2
\Bigr) \left( L_2 + L_1 \right) 
-\frac{2312}{9261} H_0(\eta ) \left( L_2 - L_1 \right)
+\left(
        \frac{78873}{219520}
        +\frac{3872}{6615} H_1(\eta )
\right) \NN \\ &&
\times H_0^2(\eta )
+\frac{1936}{19845} H_0^3(\eta )
-\frac{7744}{6615} H_0(\eta ) H_{0,1}(\eta )
+\frac{7744}{6615} H_{0,0,1}(\eta )
-\frac{15184}{99225} \zeta_2
+\frac{15488}{19845} \zeta_3 \NN \\ &&
+\left(
        \frac{27687011}{31752000}
        +\frac{342121}{3386880} H_0^2(\eta )
\right) \left( \eta + \frac{1}{\eta} \right)
+\frac{603709}{1411200} H_0(\eta ) \left( \eta - \frac{1}{\eta} \right)
+\Bigl(
        -\frac{5441}{23520} \NN \\ &&
        -\frac{5261}{188160} H_0^2(\eta )
\Bigr) \left( \eta^2 + \frac{1}{\eta^2} \right)
-\frac{1349}{11760} H_0(\eta ) \left( \eta^2 - \frac{1}{\eta^2} \right)
+\left(
        \frac{81}{3136}        
        +\frac{81}{25088} H_0^2(\eta )
\right) \NN \\ &&
\times \left( \eta^3 + \frac{1}{\eta^3} \right) 
+\frac{81}{6272} H_0(\eta ) \left( \eta^3 - \frac{1}{\eta^3} \right)
+\Biggl[
        \left(
                \frac{26939}{13440} H_{0,-1}\big(\sqrt{\eta }\big)
                +\frac{26939}{53760} H_{0,1}(\eta )
        \right) H_0(\eta ) \NN \\ &&
        -\frac{26939}{107520} \big(
                 H_{-1}\big(\sqrt{\eta }\big)
                + H_1\big(\sqrt{\eta }\big)
        \big) H_0^2(\eta )
        -\frac{26939}{6720} H_{0,0,-1}\big(\sqrt{\eta }\big)
        -\frac{26939}{53760} H_{0,0,1}(\eta )
\Biggr] \left( \sqrt{\eta} + \frac{1}{\sqrt{\eta}} \right) \NN \\ &&
+\Biggl[
        \left(
                \frac{10649}{24192} H_{0,-1}\big(\sqrt{\eta }\big)
                +\frac{10649}{96768} H_{0,1}(\eta )
        \right) H_0(\eta )
        -\frac{10649}{193536} \big(
                 H_{-1}\big(\sqrt{\eta }\big)
                + H_1\big(\sqrt{\eta }\big)
        \big) H_0^2(\eta ) \NN \\ &&
        -\frac{10649}{12096} H_{0,0,-1}\big(\sqrt{\eta }\big)
        -\frac{10649}{96768} H_{0,0,1}(\eta )
\Biggr] \left( \eta^{3/2} + \frac{1}{\eta^{3/2}} \right)
+\Biggl[
        \Bigl(
                -\frac{223}{1920} H_{0,-1}\big(\sqrt{\eta }\big) \NN \\ &&
                -\frac{223}{7680} H_{0,1}(\eta )
        \Bigr) H_0(\eta )
        +\frac{223}{15360} \big(
                 H_{-1}\big(\sqrt{\eta }\big)
                + H_1\big(\sqrt{\eta }\big)
        \big) H_0^2(\eta )
        +\frac{223}{960} H_{0,0,-1}\big(\sqrt{\eta }\big) \NN \\ &&
        +\frac{223}{7680} H_{0,0,1}(\eta )
\Biggr] \left( \eta^{5/2} + \frac{1}{\eta^{5/2}} \right)
+\Biggl[
        \left(
                \frac{81}{6272} H_{0,-1}\big(\sqrt{\eta }\big)
                +\frac{81}{25088} H_{0,1}(\eta )
        \right) H_0(\eta ) \NN \\ &&
        -\frac{81}{50176} \big(
                 H_{-1}\big(\sqrt{\eta }\big)
                + H_1\big(\sqrt{\eta }\big)
        \big) H_0^2(\eta )
        -\frac{81}{3136} H_{0,0,-1}\big(\sqrt{\eta }\big)
        -\frac{81}{25088} H_{0,0,1}(\eta )
\Biggr] \NN \\ &&
\times \left( \eta^{7/2} + \frac{1}{\eta^{7/2}} \right)
\end{eqnarray}


\begin{eqnarray}
\lefteqn{A_{Qq}^{{\sf PS},(3)}  \left( N=8 \right) =} \NN \\ && 
-\frac{87616}{25515 \ep^3}
+\frac{1}{\ep^2}
\Bigl[
\frac{4916}{107163}
-\frac{21904}{8505} \left( L_2 + L_1 \right)
\Bigr]
+\frac{1}{\ep}
\Bigl[
-\frac{33262473901}{10126903500}
-\frac{5476}{2835} \left( L_2^2 + L_1^2 \right) \NN \\ &&
+\frac{1229}{35721} \left( L_2 + L_1 \right)
-\frac{5476}{2835} H_0(\eta ) \left( L_2 - L_1 \right)
-\frac{10952}{8505} H_0^2(\eta )
-\frac{10952}{8505} \zeta_2
\Bigr]
-\frac{8273033473567}{27221116608000} \NN \\ &&
-\frac{2738}{2835} \left( L_2^3 + L_1^3 \right)
+\frac{1229}{47628} \left( L_2^2 + L_1^2 \right)
-\frac{1369}{945} H_0(\eta ) \left( L_2^2 - L_1^2 \right)
+\Bigl(
        -\frac{33262473901}{13502538000} \NN \\ &&
        -\frac{2738}{2835} H_0(\eta )^2
        -\frac{2738}{2835} \zeta_2
\Bigr) \left( L_2 + L_1 \right)
-\frac{40333}{510300} H_0(\eta ) \left( L_2 - L_1 \right)
+\Bigl(
        \frac{328686091}{1567641600} \NN \\ &&
        +\frac{2738}{8505} H_1(\eta )
\Bigr) H_0^2(\eta )
+\frac{1369}{25515} H_0^3(\eta )
-\frac{5476}{8505} H_0(\eta ) H_{0,1}(\eta )
+\frac{5476}{8505} H_{0,0,1}(\eta )
+\frac{1229}{71442} \zeta_2 \NN \\ &&
+\frac{10952}{25515} \zeta_3
+\Bigl(
        \frac{171113081}{304819200}
        +\frac{4243147}{69672960} H_0^2(\eta )
\Bigr) \left( \eta + \frac{1}{\eta} \right)
+\frac{4720627}{17418240} H_0(\eta ) \left( \eta - \frac{1}{\eta} \right) \NN \\ &&
+\Bigl(
        -\frac{30598577}{108864000}
        -\frac{1158389}{34836480} H_0^2(\eta )
\Bigr) \left( \eta^2 + \frac{1}{\eta^2} \right)
-\frac{6036587}{43545600} H_0(\eta ) \left( \eta^2 - \frac{1}{\eta^2} \right) \NN \\ &&
+\Bigl(
        \frac{2487251}{47029248}
        +\frac{271091}{41803776} H_0^2(\eta )
\Bigr) \left( \eta^3 + \frac{1}{\eta^3} \right) 
+\frac{825131}{31352832} H_0(\eta ) \left( \eta^3 - \frac{1}{\eta^3} \right) \NN \\ &&
+\Bigl(
        -\frac{847}{248832}
        -\frac{847}{1990656} H_0^2(\eta )
\Bigr) \left( \eta^4 + \frac{1}{\eta^4} \right)
-\frac{847}{497664} H_0(\eta ) \left( \eta^4 - \frac{1}{\eta^4} \right) \NN \\ &&
+\Biggl[
        \left(
                \frac{48113}{43008} H_{0,-1}\big(\sqrt{\eta }\big)
                +\frac{48113}{172032} H_{0,1}(\eta )
        \right) H_0(\eta )
        -\frac{48113}{344064} \big(
                 H_{-1}\big(\sqrt{\eta }\big)
                + H_1\big(\sqrt{\eta }\big)
        \big) H_0^2(\eta ) \NN \\ &&
        -\frac{48113}{21504} H_{0,0,-1}\big(\sqrt{\eta }\big)
        -\frac{48113}{172032} H_{0,0,1}(\eta )
\Biggr] \left( \sqrt{\eta} + \frac{1}{\sqrt{\eta}} \right)
+\Bigl(
        \Bigl(
                \frac{331775}{1161216} H_{0,-1}\big(\sqrt{\eta }\big) \NN \\ &&
                +\frac{331775}{4644864} H_{0,1}(\eta )
        \Bigr) H_0(\eta )
        -\frac{331775}{9289728} \big(
                 H_{-1}\big(\sqrt{\eta }\big)
                + H_1\big(\sqrt{\eta }\big)
        \big) H_0^2(\eta )
        -\frac{331775}{580608} H_{0,0,-1}\big(\sqrt{\eta }\big) \NN \\ &&
        -\frac{331775}{4644864} H_{0,0,1}(\eta )
\Bigr) \left( \eta^{3/2} + \frac{1}{\eta^{3/2}} \right)
+\Biggl[
        \left(
                -\frac{1449}{10240} H_{0,-1}\big(\sqrt{\eta }\big)
                -\frac{1449}{40960} H_{0,1}(\eta )
        \right) H_0(\eta ) \NN \\ &&
        +\frac{1449}{81920}\big(
                 H_{-1}\big(\sqrt{\eta }\big)
                + H_1\big(\sqrt{\eta }\big)
        \big) H_0^2(\eta )
        +\frac{1449}{5120} H_{0,0,-1}\big(\sqrt{\eta }\big)
        +\frac{1449}{40960} H_{0,0,1}(\eta )
\Biggr] \left( \eta^{5/2} + \frac{1}{\eta^{5/2}} \right) \NN \\ &&
+\Biggl[
        \left(
                \frac{95}{3584} H_{0,-1}\big(\sqrt{\eta }\big)
                +\frac{95}{14336} H_{0,1}(\eta )
        \right) H_0(\eta )
        -\frac{95}{28672} \big(
                 H_{-1}\big(\sqrt{\eta }\big)
                + H_1\big(\sqrt{\eta }\big)
        \big) H_0^2(\eta ) \NN \\ &&
        -\frac{95}{1792} H_{0,0,-1}\big(\sqrt{\eta }\big)
        -\frac{95}{14336} H_{0,0,1}(\eta )
\Biggr] \left( \eta^{7/2} + \frac{1}{\eta^{7/2}} \right)
+\Bigl(
        \Bigl(
                -\frac{847}{497664} H_{0,-1}\big(\sqrt{\eta }\big) \NN \\ &&
                -\frac{847}{1990656} H_{0,1}(\eta )
        \Bigr) H_0(\eta )
        + \frac{847}{3981312} \big(
                 H_{-1}\big(\sqrt{\eta }\big)
                + H_1\big(\sqrt{\eta }\big)
        \big) H_0^2(\eta )
        +\frac{847}{248832} H_{0,0,-1}\big(\sqrt{\eta }\big) \NN \\ &&
        +\frac{847}{1990656} H_{0,0,1}(\eta )
\Bigr) \left( \eta^{9/2} + \frac{1}{\eta^{9/2}} \right)
\end{eqnarray}


\begin{eqnarray}
\lefteqn{A_{Qq}^{{\sf PS},(3)} \left( N=10 \right) =} \NN \\ && 
-\frac{1605632}{735075\ep^3}
+\frac{1}{\ep^2}
\Bigl[
\frac{5105152}{33078375}
-\frac{401408}{245025} \left( L_2 + L_1 \right) 
\Bigr]
+\frac{1}{\ep}
\Bigl[
-\frac{2689775322848}{1260782263125}
-\frac{100352}{81675} \left( L_2^2 + L_1^2 \right) \NN \\ &&
+\frac{1276288}{11026125} \left( L_2 + L_1 \right)
-\frac{100352}{81675} H_0(\eta ) \left( L_2 - L_1 \right)
-\frac{200704}{245025} H_0^2(\eta )
-\frac{200704}{245025} \zeta_2
\Bigr] \NN \\ &&
-\frac{19054928458130951}{406677926793600000}
-\frac{50176}{81675} \left( L_2^3 + L_1^3 \right)
+\frac{319072}{3675375} \left( L_2^2 + L_1^2 \right)
-\frac{25088}{27225} H_0(\eta ) \left( L_2^2 - L_1^2 \right) \NN \\ &&
+\Bigl(
        -\frac{672443830712}{420260754375}
        -\frac{50176}{81675} H_0^2(\eta )
        -\frac{50176}{81675} \zeta_2
\Bigr) \left( L_2 + L_1 \right)
-\frac{436544}{13476375} H_0(\eta ) \left( L_2 - L_1 \right) \NN \\ &&
+\Bigl(
        \frac{43556878529}{331195392000}
        +\frac{50176}{245025} H_1(\eta )
\Bigr) H_0^2(\eta )
+\frac{25088}{735075} H_0^3(\eta )
-\frac{100352}{245025} H_0(\eta ) H_{0,1}(\eta ) \NN \\ &&
+\frac{100352}{245025} H_{0,0,1}(\eta )
+\frac{638144}{11026125} \zeta_2
+\frac{200704}{735075} \zeta_3
+\Bigl(
        \frac{226878798767}{526901760000}
        +\frac{347257523}{8028979200} H_0^2(\eta )
\Bigr) \NN \\ &&
\times \left( \eta + \frac{1}{\eta} \right)
+\frac{2047449637}{10036224000} H_0(\eta ) \left( \eta - \frac{1}{\eta} \right)
+\Bigl(
        -\frac{368396509553}{1075757760000}
        -\frac{55227289}{1405071360} H_0^2(\eta )
\Bigr) \NN \\ &&
\times  \left( \eta^2 + \frac{1}{\eta^2} \right)
-\frac{41219216111}{245887488000} H_0(\eta ) \left( \eta^2 - \frac{1}{\eta^2} \right)
+\Bigl(
        \frac{27528100609}{270978048000}
        +\frac{118201777}{9634775040} H_0^2(\eta )
\Bigr) \NN \\ &&
\times \left( \eta^3 + \frac{1}{\eta^3} \right)
+\frac{1819513853}{36130406400} H_0(\eta ) \left( \eta^3 - \frac{1}{\eta^3} \right)
+\Bigl(
        -\frac{1197239}{100362240}
        -\frac{1182029}{802897920} H_0^2(\eta )
\Bigr) \NN \\ &&
\times \left( \eta^4 + \frac{1}{\eta^4} \right)
-\frac{2386873}{401448960} H_0(\eta ) \left( \eta^4 - \frac{1}{\eta^4} \right)
+\Bigl(
        \frac{507}{991232}
        +\frac{507}{7929856} H_0^2(\eta )
\Bigr) \left( \eta^5 + \frac{1}{\eta^5} \right) \NN \\ &&
+\frac{507}{1982464} H_0(\eta ) \left( \eta^5 - \frac{1}{\eta^5} \right)
+\Biggl[
        \left(
                \frac{980747}{1351680} H_{0,-1}\big(\sqrt{\eta }\big)
                +\frac{980747}{5406720}  H_{0,1}(\eta )
        \right) H_0(\eta ) \NN \\ &&
        -\frac{980747}{10813440} \big(
                 H_{-1}\big(\sqrt{\eta }\big)
                + H_1\big(\sqrt{\eta }\big)
        \big) H_0^2(\eta )
        -\frac{980747}{675840} H_{0,0,-1}\big(\sqrt{\eta }\big)
        -\frac{980747}{5406720} H_{0,0,1}(\eta )
\Biggr] \NN \\ &&
\times \left( \sqrt{\eta} + \frac{1}{\sqrt{\eta}} \right)
+\Biggl[
        \left(
                \frac{734267}{3317760} H_{0,-1}\big(\sqrt{\eta }\big)
                +\frac{734267}{13271040} H_{0,1}(\eta )
        \right) H_0(\eta )
        -\frac{734267}{26542080} \NN \\ &&
\times \big(
                 H_{-1}\big(\sqrt{\eta }\big)
                + H_1\big(\sqrt{\eta }\big)
        \big) H_0^2(\eta )
        -\frac{734267}{1658880} H_{0,0,-1}\big(\sqrt{\eta }\big)
        -\frac{734267}{13271040} H_{0,0,1}(\eta )
\Biggr] \NN \\ &&
\times \left( \eta^{3/2} + \frac{1}{\eta^{3/2}} \right)
+\Biggl[
        \left(
                -\frac{70889}{409600} H_{0,-1}\big(\sqrt{\eta }\big)
                -\frac{70889}{1638400} H_{0,1}(\eta )
        \right) H_0(\eta ) 
        +\frac{70889}{3276800} \NN \\ && 
\times \big(
                 H_{-1}\big(\sqrt{\eta }\big)
                + H_1\big(\sqrt{\eta }\big)
        \big) H_0^2(\eta )
        +\frac{70889}{204800} H_{0,0,-1}\big(\sqrt{\eta }\big)
        +\frac{70889}{1638400} H_{0,0,1}(\eta )
\Biggr] \NN \\ && 
\times \left( \eta^{5/2} + \frac{1}{\eta^{5/2}} \right)
+\Biggl[
        \left(
                \frac{4179}{81920} H_{0,-1}\big(\sqrt{\eta }\big)
                +\frac{4179}{327680} H_{0,1}(\eta )
        \right) H_0(\eta ) \NN \\ &&
        -\frac{4179}{655360} \big(
                 H_{-1}\big(\sqrt{\eta }\big)
                + H_1\big(\sqrt{\eta }\big)
        \big) H_0^2(\eta )
        -\frac{4179}{40960} H_{0,0,-1}\big(\sqrt{\eta }\big)
        -\frac{4179}{327680} H_{0,0,1}(\eta )
\Biggr] \NN \\ &&
\times \left( \eta^{7/2} + \frac{1}{\eta^{7/2}} \right)
+\Biggl[
        \left(
                -\frac{39641}{6635520} H_{0,-1}\big(\sqrt{\eta }\big)
                -\frac{39641}{26542080} H_{0,1}(\eta )
        \right) H_0(\eta ) \NN \\ &&
        +\frac{39641}{53084160} \big(
                 H_{-1}\big(\sqrt{\eta }\big)
                + H_1\big(\sqrt{\eta }\big)
        \big) H_0^2(\eta )
        +\frac{39641}{3317760} H_{0,0,-1}\big(\sqrt{\eta }\big)
        +\frac{39641}{26542080} H_{0,0,1}(\eta )
\Biggr] \NN \\ &&
\times \left( \eta^{9/2} + \frac{1}{\eta^{9/2}} \right)
+\Biggl[
        \left(
                \frac{507}{1982464} H_{0,-1}\big(\sqrt{\eta }\big)
                +\frac{507}{7929856} H_{0,1}(\eta )
        \right) H_0(\eta ) 
        -\frac{507}{15859712} \NN \\ &&
\times \big(
                 H_{-1}\big(\sqrt{\eta }\big)
                + H_1\big(\sqrt{\eta }\big)
        \big) H_0^2(\eta )
        -\frac{507}{991232} H_{0,0,-1}\big(\sqrt{\eta }\big)
        -\frac{507}{7929856} H_{0,0,1}(\eta )
\Biggr] \NN \\ &&
\times \left( \eta^{11/2} + \frac{1}{\eta^{11/2}} \right).
\end{eqnarray}
The above expressions depend on $\eta$ only, but not on $\sqrt{\eta}$, and they are symmetric for $\eta 
\leftrightarrow \eta^{-1}$. The expansions of the the $O(\varepsilon^0)$ 
terms for $N = 2,4,6$  up to $O(\eta^3 \ln^3(\eta))$ for $\eta < 1$ agree with the results given in 
Ref.~\cite{Ablinger:2017err}.
With growing values of $N$, these expressions exhibit a growing degree of the polynomials in $\eta$. We have 
found that the general $N$ formula cannot be expressed as a sum--product solution by means of difference 
field theory. This means that the corresponding solution will be given by a higher transcendental function
depending on $N$ and $\eta$. We will study this behaviour elsewhere.

\vspace*{5mm}
\noindent
{\bf Acknowledgment.}\\
This work
was supported in part by the Austrian Science Fund (FWF) grant SFB F50 (F5009-N15), the European
Commission through contract PITN-GA-2012-316704 ({HIGGSTOOLS}).



\begin{thebibliography}{99}
%
\bibitem{VFNS}
J.~Bl\"umlein, A.~De Freitas, C.~Schneider, and K.~Sch\"onwald, DESY 17-187.
%
\bibitem{Ablinger:2017err}
  J.~Ablinger, J.~Bl\"umlein, A.~De Freitas, A.~Hasselhuhn, C.~Schneider and F.~Wi{\ss}brock,
  Nucl.\ Phys.\ B {\bf 921} (2017) 585
  [arXiv:1705.07030[hep-ph]].
%
\bibitem{Ablinger:2014nga}
  J.~Ablinger, A.~Behring, J.~Bl\"umlein, A.~De Freitas, A.~von Manteuffel and C.~Schneider,
  Nucl.\ Phys.\ B {\bf 890} (2014) 48
  [arXiv:1409.1135 [hep-ph]].
%
\bibitem{MOM}
  J.~Ablinger, J.~Bl\"umlein, S.~Klein, C.~Schneider and F.~Wi{\ss}brock,
  arXiv:1106.5937 [hep-ph];\\
  J.~Ablinger, J.~Bl\"umlein, A.~Hasselhuhn, S.~Klein, C.~Schneider and F.~Wi{\ss}brock,
  PoS RADCOR {\bf 2011} (2011) 031
  [arXiv:1202.2700 [hep-ph]].
%
\bibitem{Harlander:1997zb}
  R.~Harlander, T.~Seidensticker and M.~Steinhauser,
  Phys.\ Lett.\ B {\bf 426} (1998) 125
  [hep-ph/9712228].
%
\bibitem{Seidensticker:1999bb}
  T.~Seidensticker,
  hep-ph/9905298.
%
\bibitem{AGG2}
J.~Ablinger et al., in preparation.
%
\bibitem{Ablinger:2014lka}
  J.~Ablinger, J.~Bl\"umlein, A.~De Freitas, A.~Hasselhuhn, A.~von Manteuffel, M.~Round, C.~Schneider and F.~Wi{\ss}brock,
  Nucl.\ Phys.\ B {\bf 882} (2014) 263
  [arXiv:1402.0359 [hep-ph]].
%
\bibitem{Ablinger:2014uka}
  J.~Ablinger, J.~Bl\"umlein, A.~De Freitas, A.~Hasselhuhn, A.~von Manteuffel, M.~Round and C.~Schneider,
  Nucl.\ Phys.\ B {\bf 885} (2014) 280
  [arXiv:1405.4259 [hep-ph]].
%
\bibitem{Ablinger:2014vwa}
J.~Ablinger, A.~Behring, J.~Bl\"umlein, A.~De Freitas, A. Hasselhuhn, A.~von Manteuffel, M.~Round, 
C.~Schneider, and F.~Wi{\ss}brock,
  Nucl.\ Phys.\ B {\bf 886} (2014) 733
  [arXiv:1406.4654 [hep-ph]].
%
\bibitem{Blumlein:2016xcy}
  J.~Bl\"umlein, G.~Falcioni and A.~De Freitas,
  Nucl.\ Phys.\ B {\bf 910} (2016) 568
  [arXiv:1605.05541 [hep-ph]].
%
\bibitem{Behring:2016hpa}
  A.~Behring, J.~Bl\"umlein, G.~Falcioni, A.~De Freitas, A.~von Manteuffel and C.~Schneider,
  Phys.\ Rev.\ D {\bf 94} (2016) no.11,  114006
  [arXiv:1609.06255 [hep-ph]].
%
\bibitem{Ablinger:2010ty}
  J.~Ablinger, J.~Bl\"umlein, S.~Klein, C.~Schneider and F.~Wi{\ss}brock,
  Nucl.\ Phys.\ B {\bf 844} (2011) 26
  [arXiv:1008.3347 [hep-ph]].
%
\bibitem{Blumlein:2012vq}
  J.~Bl\"umlein, A.~Hasselhuhn, S.~Klein and C.~Schneider,
  Nucl.\ Phys.\ B {\bf 866} (2013) 196
  [arXiv:1205.4184 [hep-ph]].
%
\bibitem{Behring:2015zaa}
  A.~Behring, J.~Bl\"umlein, A.~De Freitas, A.~von Manteuffel and C.~Schneider,
  Nucl.\ Phys.\ B {\bf 897} (2015) 612
  [arXiv:1504.08217 [hep-ph]].
%
\bibitem{Behring:2015roa}
  A.~Behring, J.~Bl\"umlein, A.~De Freitas, A.~Hasselhuhn, A.~von Manteuffel and C.~Schneider,
  Phys.\ Rev.\ D {\bf 92} (2015) no.11,  114005
  [arXiv:1508.01449 [hep-ph]].
%
\bibitem{AGG}
J.~Ablinger et al., DESY 15--112.
%
\bibitem{Behring:2014eya}
  A.~Behring, I.~Bierenbaum, J.~Bl\"umlein, A.~De Freitas, S.~Klein and F.~Wi{\ss}brock,
  Eur.\ Phys.\ J.\ C {\bf 74} (2014) no.9,  3033
  [arXiv:1403.6356 [hep-ph]].
%
\bibitem{Ablinger:2016swq}
J.~Ablinger, A.~Behring, J.~Bl\"umlein, G.~Falcioni, A.~De Freitas,
A.~Hasselhuhn, A.~von Manteuffel, M.~Round, C. Schneider, and F.~Wi{\ss}brock,
  PoS LL {\bf 2016} (2016) 065
  [arXiv:1609.03397 [hep-ph]].
%
\bibitem{Ablinger:2015tua}
  J.~Ablinger, A.~Behring, J.~Bl\"umlein, A.~De Freitas, A.~von Manteuffel and C.~Schneider,
  Comput.\ Phys.\ Commun.\  {\bf 202} (2016) 33
  [arXiv:1509.08324 [hep-ph]].
%
\bibitem{PROC1}
J.~Ablinger, A.~Behring, J.~Bl\"umlein, A.~De Freitas, A.~von Manteuffel, and
       C.~Schneider, DESY 17-199.

%
\bibitem{Bierenbaum:2009mv}
  I.~Bierenbaum, J.~Bl\"umlein and S.~Klein,
  Nucl.\ Phys.\ B {\bf 820} (2009) 417
  [arXiv:0904.3563 [hep-ph]].
%
\bibitem{Steinhauser:2000ry}
  M.~Steinhauser,
  Comput.\ Phys.\ Commun.\  {\bf 134} (2001) 335
  [hep-ph/0009029].
%
\bibitem{Ablinger:2017tan}
  J.~Ablinger, A.~Behring, J.~Bl\"umlein, A.~De Freitas, A.~von Manteuffel and C.~Schneider,
  Nucl.\ Phys.\ B {\bf 922} (2017) 1
  [arXiv:1705.01508 [hep-ph]].
%
\bibitem{Blumlein:2011mi}
  J.~Bl\"umlein, A.~De Freitas and W.~van Neerven,
  Nucl.\ Phys.\ B {\bf 855} (2012) 508
  [arXiv:1107.4638 [hep-ph]].
%
\bibitem{Buza:1995ie}
  M.~Buza, Y.~Matiounine, J.~Smith, R.~Migneron and W.L.~van Neerven,
  Nucl.\ Phys.\ B {\bf 472} (1996) 611
  [hep-ph/9601302].
%
\bibitem{Buza:1996wv}
  M.~Buza, Y.~Matiounine, J.~Smith and W.L.~van Neerven,
  Eur.\ Phys.\ J.\ C {\bf 1} (1998) 301
  [hep-ph/9612398].
%
\bibitem{Bierenbaum:2007qe}
  I.~Bierenbaum, J.~Bl\"umlein and S.~Klein,
  Nucl.\ Phys.\ B {\bf 780} (2007) 40
  [hep-ph/0703285].
%
\bibitem{Bierenbaum:2008yu}
  I.~Bierenbaum, J.~Bl\"umlein, S.~Klein and C.~Schneider,
  Nucl.\ Phys.\ B {\bf 803} (2008) 1
  [arXiv:0803.0273 [hep-ph]].
%
\bibitem{Bierenbaum:2009zt}
  I.~Bierenbaum, J.~Bl\"umlein and S.~Klein,
  Phys.\ Lett.\ B {\bf 672} (2009) 401
  [arXiv:0901.0669 [hep-ph]].
%
\bibitem{MB1a}
E.W.~Barnes, 
Proc. Lond. Math. Soc. (2) {\bf 6} (1908) 141.
%
\bibitem{MB1b}
E.W.~Barnes,
Quarterly Journal of Mathematics {\bf 41} (1910) 136.
%
\bibitem{MB2}
H.~Mellin,
Math. Ann. {\bf 68}, no. 3 (1910) 305.
%
\bibitem{MB3}
E.T.~Whittaker and G.N.~Watson, {\sf A Course of Modern Analysis}, (Cambridge University Press, Cambridge, 1927;
                   reprinted 1996). 
%
\bibitem{MB4}
E.C.~Titchmarsh,
{\sf Introduction to the Theory of Fourier Integrals},
(Calendron Press, Oxford, 1937; 2nd Edition 1948).
%
\bibitem{FORM}
  J.A.M.~Vermaseren,
  math-ph/0010025.
%
\bibitem{MB}
  M.~Czakon,
  Comput.\ Phys.\ Commun.\  {\bf 175} (2006) 559
  [hep-ph/0511200].
%
\bibitem{MBr}
  A.V.~Smirnov and V.A.~Smirnov,
  Eur.\ Phys.\ J.\ C {\bf 62} (2009) 445
  [arXiv:0901.0386 [hep-ph]].
%
\bibitem{HSUM}
  J.A.M.~Vermaseren,
  Int.\ J.\ Mod.\ Phys.\ A {\bf 14} (1999) 2037
  [hep-ph/9806280];\\
  J.~Bl\"umlein and S.~Kurth,
  Phys.\ Rev.\  D {\bf 60} (1999) 014018
  [arXiv:hep-ph/9810241].
%
\bibitem{SIG1}
C.~Schneider, {S\'em.~Lothar. Combin.\/} {\bf 56} (2007) 1, 
 article B56b.
%
\bibitem{SIG2}
C.~Schneider, {{\sf Computer Algebra in Quantum Field Theory: Integration,
  Summation and Special Functions}\/} Texts and Monographs in Symbolic
  Computation eds. C.~Schneider and J.~Bl\"umlein  (Springer, Wien, 2013) 325, 
  arXiv:1304.4134 [cs.SC].
%
\bibitem{HARMONICSUMS}
  J.~Ablinger,
  PoS {(LL2014)} 019;
  {\sf Computer Algebra Algorithms for Special Functions in Particle Physics}, Ph.D. Thesis, J. Kepler University 
Linz, 2012,
  arXiv:1305.0687 [math-ph];\\
  {\sf A Computer Algebra Toolbox for Harmonic Sums Related to Particle Physics}, Diploma Thesis, J. Kepler University Linz, 2009,
  arXiv:1011.1176 [math-ph].
%
\bibitem{Ablinger:2011te}
  J.~Ablinger, J.~Bl\"umlein and C.~Schneider,
  J.\ Math.\ Phys.\  {\bf 52} (2011) 102301
  [arXiv:1105.6063 [math-ph]].
%
\bibitem{Ablinger:2013cf}
  J.~Ablinger, J.~Bl\"umlein and C.~Schneider,
  J.\ Math.\ Phys.\  {\bf 54} (2013) 082301
  [arXiv:1302.0378 [math-ph]].
%
\bibitem{EMSSP}
  J.~Ablinger, J.~Bl\"umlein, S.~Klein and C.~Schneider,
  Nucl.\ Phys.\ Proc.\ Suppl.\  {\bf 205-206} (2010) 110
  [arXiv:1006.4797 [math-ph]];\\
  J.~Bl\"umlein, A.~Hasselhuhn and C.~Schneider,
  PoS (RADCOR 2011) 032
  [arXiv:1202.4303 [math-ph]];\\
  C.~Schneider,
  J.\ Phys.\ Conf.\ Ser.\  {\bf 523} (2014) 012037
  [arXiv:1310.0160 [cs.SC]].
%
\bibitem{Remiddi:1999ew}
  E.~Remiddi and J.~A.~M.~Vermaseren,
  Int.\ J.\ Mod.\ Phys.\ A {\bf 15} (2000) 725
  [hep-ph/9905237].
%
\bibitem{Ablinger:2014bra}
  J.~Ablinger, J.~Bl\"umlein, C.G.~Raab and C.~Schneider,
  J.\ Math.\ Phys.\  {\bf 55} (2014) 112301
  [arXiv:1407.1822 [hep-th]].
%
\bibitem{Blumlein:2003gb}
  J.~Bl\"umlein,
  Comput.\ Phys.\ Commun.\  {\bf 159} (2004) 19
  [hep-ph/0311046].
%
\bibitem{Alekhin:2012vu}
  S.~Alekhin, J.~Bl\"umlein, K.~Daum, K.~Lipka and S.~Moch,
  Phys.\ Lett.\ B {\bf 720} (2013) 172
  [arXiv:1212.2355 [hep-ph]].
%
\bibitem{Agashe:2014kda}
  K.A.~Olive {\it et al.} [Particle Data Group],
  Chin.\ Phys.\ C {\bf 38} (2014) 090001.
%
\bibitem{Blumlein:2017dxp}
  J.~Bl\"umlein and C.~Schneider,
  Phys.\ Lett.\ B {\bf 771} (2017) 31
  [arXiv:1701.04614 [hep-ph]].
%
\bibitem{LEWIN1}
L.~Lewin, {\sf Dilogarithms and Associated Functions}, (Macdonald, London, 1958).
%
\bibitem{LEWIN2}
L.~Lewin, {\sf Polylogarithms and Associated Functions}, (North-Holland, New York, 
1981).
\end{thebibliography}
\end{document}